\documentclass[twocolumn,iop,revtex4]{openjournal}
\usepackage{natbib} 
\usepackage[dvipsnames]{xcolor}
\usepackage{aas_macros} 
\usepackage{amssymb}
\usepackage{amsmath}
\usepackage[title]{appendix}
\usepackage{hyperref}	
\hypersetup{colorlinks=true,linkcolor=blue,citecolor=blue,filecolor=blue,urlcolor=blue}
\usepackage[caption=false]{subfig}
\usepackage{soul}                          
\usepackage[raggedrightboxes]{ragged2e}
\newcommand{\lya}{Lyman-$\alpha$~}

\newcommand{\enzo}{\texttt{Enzo~}}

\newcommand{\yt}{\texttt{yt~}}

\newcommand{\grackle}{\texttt{Grackle-2.1~}}

\newcommand{\kms} {km $\rm{s^{-1}}$}
\newcommand{\mpch} {$\rm{h^{-1}}$ Mpc\,\,}
\newcommand{\mpchv} {\rm $h^{-3}$ Mpc$^3$\,\,} 
\newcommand{\kpch} {\rm $h^{-1}$ kpc\,\,} 
\newcommand{\msolar} {$\rm{M_{\odot}}~$}
\newcommand{\msolarc} {$\rm{M_{\odot}}$}
\newcommand{\msolaryr} {$\rm{M_{\odot}~yr^{-1}}~$}
\newcommand{\msolaryrc} {$\rm{M_{\odot}~yr^{-1}}$}

\newcommand{\zsolar} {$\rm{Z_{\odot}}~$}

\newcommand{\molH} {$\rm{H_2}$~}
\newcommand{\molHc} {$\rm{H_2}$}

\newcommand{\blackdemon} {\texttt{BlackDemon~}}
\newcommand{\renaissance} {\texttt{Renaissance~}}
\newcommand{\rockstar} {\texttt{Rockstar~}}

\newcommand{\change}[2][]{%
\ifthenelse{\isempty{#2}}{{\color{ForestGreen}{#1}}}%
{{\color{RedOrange}\sout{#1}}{\color{ForestGreen}{ #2}}}%
}

\begin{document}
\title[The First Stars were Super Massive]{Massive Star Formation in Overdense Regions of the Early Universe}
\author{John A. Regan$^{*}$}
\thanks{$^*$E-mail:john.regan@mu.ie, Royal Society - SFI University Research Fellow}

\affiliation{Centre for Astrophysics and Space Science Maynooth, Department of Theoretical Physics, Maynooth University, Maynooth, Ireland}

\begin{abstract}
  Both the origin of, and the population demographics of, massive black holes (MBHs) remains an open
  question in modern day astrophysics. Here we introduce the \texttt{BlackDemon} suite of
  cosmological simulations using the \enzo code. The suite consists primarily of three, high resolution,
  distinct regions, each with a side length of 1 {\rm $h^{-1}$ Mpc}. Two of the regions evolve within a larger
  overdense region while the other evolves within a more `normal' region.
  The simulation suite has spatial and mass resolution capable of resolving the
  formation of the first galaxies and MBHs within each region. We report here, as the first in a series
  of papers, the evolution of the simulation suite up to the point where star formation has commenced
  in each region and for 2 Myr after the onset of star formation.
  Within these environments the masses of the first objects to form have
  masses between approximately 100 \msolar and $10^4$ \msolarc. The larger mass objects form due to
  both major mergers, which trigger rapid mass inflow to the centre of the halo, and also through
  multiple minor mergers which allows the host halo to grow to close to the atomic cooling threshold. 
  In both scenarios the initially very high accretion rates quickly grow the objects to close
  to $10^4$ \msolarc.  However, accretion halts after less than 50,000 years due to gas starvation.
  The final fate of these objects in terms of fragmentation and subsequent fragment
    mergers cannot be deduced at our current resolution. In the case where fragmentation is mild such objects are likely to
  form super-massive stars before contracting to the main sequence evolving into massive population III stars and
  subsequently MBHs.
  
\end{abstract}

\section{Introduction} \label{Sec:Introduction}
\noindent Super-Massive Black Holes (SMBHs) with masses in excess of $10^6$ \msolar exist at the centre of
most massive galaxies \citep{Kormendy_2013}. Moreover, high-z observations of distant quasars
confirm that SMBHs with masses up to $10^9$ \msolar are in place less than one billion years after the
Big Bang \citep[e.g.][]{Fan_06, Venemans_2015, Matsuoka_2019, Wang_2021}. Below the supermassive
mass scale at masses between approximately $10^3$ and $10^6$ \msolar there is speculated to exist a second
population of massive black holes\footnote{I will use the generic term massive black holes to describe all
black holes masses in excess of $10^3$ \msolar in this paper} - sometimes dubbed
\textit{intermediate mass black holes}. This
intermediate mass range is key to understanding the astrophysical evolution of the entire MBH
population. \\
\indent Below this mass
scale exist the stellar mass black holes - detectable in the local universe via both electromagnetic
observations \citep[e.g.][]{Done_2007} and through gravitational wave emission \citep[e.g.][]{GW150914}. Stellar mass black holes
are abundant in the local universe (and presumably across the entire universe) as they originate
from the end-point of massive stars. The intermediate range of MBHs ($10^{3-6}$
\msolarc) have proven very challenging to detect. This is because the accretion rate onto a black
hole scales linearly with the mass of the black hole according to the Eddington formulation:
\begin{equation}
  \rm{\dot{M}_{MBH} = \frac{4 \pi G M_{MBH} m_p}{\epsilon \sigma_T c}}
\end{equation}
where $\rm{M_{MBH}}$ is the black hole mass, $\rm{m_p}$ is the proton mass, $\epsilon$ is the radiative efficiency,
$\rm{\sigma_T}$ is the Thomson cross section and $c$ is the speed of light.
  Lower mass black holes therefore accrete at significantly lower rates compared to their
more massive counterparts. The luminosity that is detectable
from these intermediate sources is therefore lower as well making their overall detection more
challenging with current instruments. Related to this is the fact that many of these intermediate mass
black holes may not necessarily reside at the galactic centre and may instead be located at
off-nuclear locations again increasing the challenge \citep{Bellovary_2010, Chiou_2018,
  Reines_2020, Druschke_2020, Beckmann_2022}. Nonetheless, over the
past decade significant progress has been made in detecting black holes
with masses below $10^6$ \msolar \citep[e.g.][]{Baldassare_2020}.\\
\indent In any case understanding the origin of all black holes within the MBH mass window is paramount.
A key challenge is then to understand what the seed masses of the MBH mass spectrum are. Does the
entire MBH mass spectrum originate from stellar mass black holes with characteristic masses
of approximately 40 \msolar which grow through accretion and mergers? Or is another channel
required to populate the MBH mass spectrum? The simplest case is to assume that the seeds
are ``light'' and that they originated from the remnants of the first stars \citep[e.g.][]{Madau_2001}. While
this is indeed the simplest explanation there are significant challenges in growing
these light seeds in order to explain the entire MBH mass spectrum. The problem is particularly acute
when attempting to use light seed growth to explain the appearance of SMBHs at $z \gtrsim 7$. In
order to achieve such growth the seed black holes would need to grow, uninterrupted at the Eddington
rate. Such a scenario has been shown to be exceedingly unlikely with numerous cosmological simulations
showing that light seeds do not grow efficiently and tend to remain at their seed masses across cosmic
time \citep{Alvarez_2009, Milosavljevic_2009, Smith_2018, Spinoso_2022}. Although see \cite{Zubovas_2021} who
argue that, though unlikely and rare, episodes of chaotic accretion may result in the growth of
light seeds to SMBH masses. Another possible avenue that may circumvent this growth hurdle (and may also apply to
heavier seeds) is the super-Eddington accretion scenario \citep[e.g.][]{Sadowski_2009, Sadowski_2014, Jiang_2017, Regan_2019}
and where, in some cases, if the gas inflow is sufficiently high a steady state can be reached
without any time-dependent oscillations \citep{Inayoshi_2016, Inayoshi_2018, Takeo_2020}. If the black hole can
find itself at the centre of an extremely dense gas cloud then this mechanism may be viable and could in principle
grow the embryonic black hole by several orders of magnitude in mass within a short timeframe.\\
\indent Alternative routes to MBH formation may be through either a dynamical runaway or via the
formation of a supermassive star. Both of these mechanisms create a ``heavy'' seed with masses
in excess of approximately $10^3$ \msolarc. Numerous pathways to a MBH have been investigated via
a dynamical pathway. Seminal work by \cite{PortegiesZwart_2004} showed that MBHs can be formed through runaway
collisions of stars in dense young star clusters. Additional research in this direction by other
authors \citep{Devecchi_2008, Katz_2015, Rizzuto_2021, Gonzalez_2021} has also shown that MBHs can form via this mechanism
albeit it is challenging to grow black holes to masses greater than $10^3$ \msolar through this process. Related
mechanisms where initial light seed black holes grow inside a nuclear star cluster
\citep{Alexander_2014, Lupi_2014, Stone_2017, Natarajan_2020, Fragione_2022}
or in other high density environments like a circumbinary disks \citep{Lupi_2016}
offer additional, promising, routes to MBHs. \\
\indent In addition to the dynamical pathway heavy seeds can form via the formation of a
very massive or super-massive star (SMS). SMSs are defined by their extremely high accretion
rate onto the stellar surface ($\dot{M} \gtrsim 0.001$ \msolaryrc)
which, through the accretion of high entropy gas, causes the photosphere
of the star to expand (so long at the accretion rate is maintained) \citep{Haemmerle_2018}. Such stars are expected to the
very red, in contrast to massive PopIII stars whose spectrum would be blue, and cool with effective
temperatures of approximately $\rm{T_{eff}} \sim 5000$ K \citep{Hosokawa_2012, Hosokawa_2013, Haemmerle_2018, Woods_2021}.
With maximum masses of greater than $10^5$ \msolar \citep{Woods_2017} these SMS stars would be
ideal progenitors for MBHs. Over the last decade many groups have attempted to model the formation
of SMSs in the first atomic cooling haloes. Atomic cooling haloes are believed to be necessary for
SMS formation as the mass inflow rate scales with the mass of the host halo. \\
\indent The formation of a SMS additionally requires metal-free (or near metal-free) gas and so
previous generations of star formation in the halo (or the halo progenitors must be avoided).
Strong, local sources of Lyman-Werner (LW) radiation can dissociate \molH thus preventing or
delaying the normal formation of PopIII stars \citep{Machacek_2001,
  Haiman_2006, Shang_2010, WolcottGreen_2011, Visbal_2014, Visbal_2014c, Visbal_2014b, Regan_2017, WolcottGreen_2019, Skinner_2020,
Schauer_2021, Kulkarni_2021}.
However, the required flux from LW radiation has been shown to be large
\citep{Regan_2014a, Latif_2014b, Latif_2014a, Regan_2018b} and metal pollution from nearby haloes
must also be avoided \citep{Agarwal_2017b} meaning that the LW channel may be difficult to achieve
in practice and may only be able to seed a subset of SMBHs. Finally, LW escape fractions may also be
  impeded from escaping early galaxies thus further reducing the effectiveness of this channel
  \citep{Kitayama_2004, Schauer_2015, Schauer_2017}.\\
\indent Baryonic streaming velocities \citep{Tseliakhovich_2010} and the
impact they have on halo formation has also been investigated as a potential channel for heavy seed MBH seed
formation. In this case the extra kinetic energy in the gas, relative to the dark matter, delays the
virialisation of the halo and therefore allows more massive haloes to develop before star
formation can be achieved. Similar to the LW pathway this can lead to ideal environmental conditions for
MBH seed formation to take place via SMS formation
\citep{Tanaka_2014, Latif_2014c, Hirano_2017, Schauer_2017}. Whether this channel can seed all MBHs is
unclear but it may be possible to seed the massive end of the MBH mass spectrum in this way \citep{Kulkarni_2021}. Finally,
the rapid assembly of haloes has also been shown to allow for SMS formation \citep{Wise_2019}. In this
case a halo grows at a sufficient rate that radiative cooling, due to \molH emission lines, cannot
overcome the dynamical heating effect of the rapid assembly \citep{Yoshida_2003a, Fernandez_2014}.
In this case the halo continues to grow without forming stars until either the growth rate slows or
the halo begins cooling via atomic line emission cooling \citep{Wise_2019}. The rapid assembly scenario
has the advantage that it may produce a higher number density of MBH seeds \citep{Regan_2020, Lupi_2021}
compared to either the LW pathway or the baryonic streaming pathway but it is still unclear whether
very high final masses can be achieved for the seeds \citep{Regan_2020b, Latif_2022}. The goal of this simulation
suite is to probe the different pathways, with the exception of the streaming velocities pathway, as self-consistently
as possible. The rapid assembly process is captured
due to the overdense nature of the environment with light seeds forming organically and captured via our star formation
prescriptions. The LW pathway will be captured if such a condition can arise in our volume. The capture of dynamical pathways
(i.e. the formation of nuclear stellar clusters) is also possible to capture as well but would require zoom-in re-simulations
to probe in detail. \\
\indent The paper is laid out as follows: In \S \ref{Sec:Methods} we describe the methods used in designing and running
  the simulations and the rationale behind them. In \S
  \ref{Sec:Results} we analyse the results of the simulations up to the point where they are now.
  In \S \ref{Sec:Discussion} we summarize our results and outline our conclusions.

\begin{figure*}
\centering
\begin{minipage}{175mm}      \begin{center}
\centerline{
    \includegraphics[width=18.0cm, height=8cm]{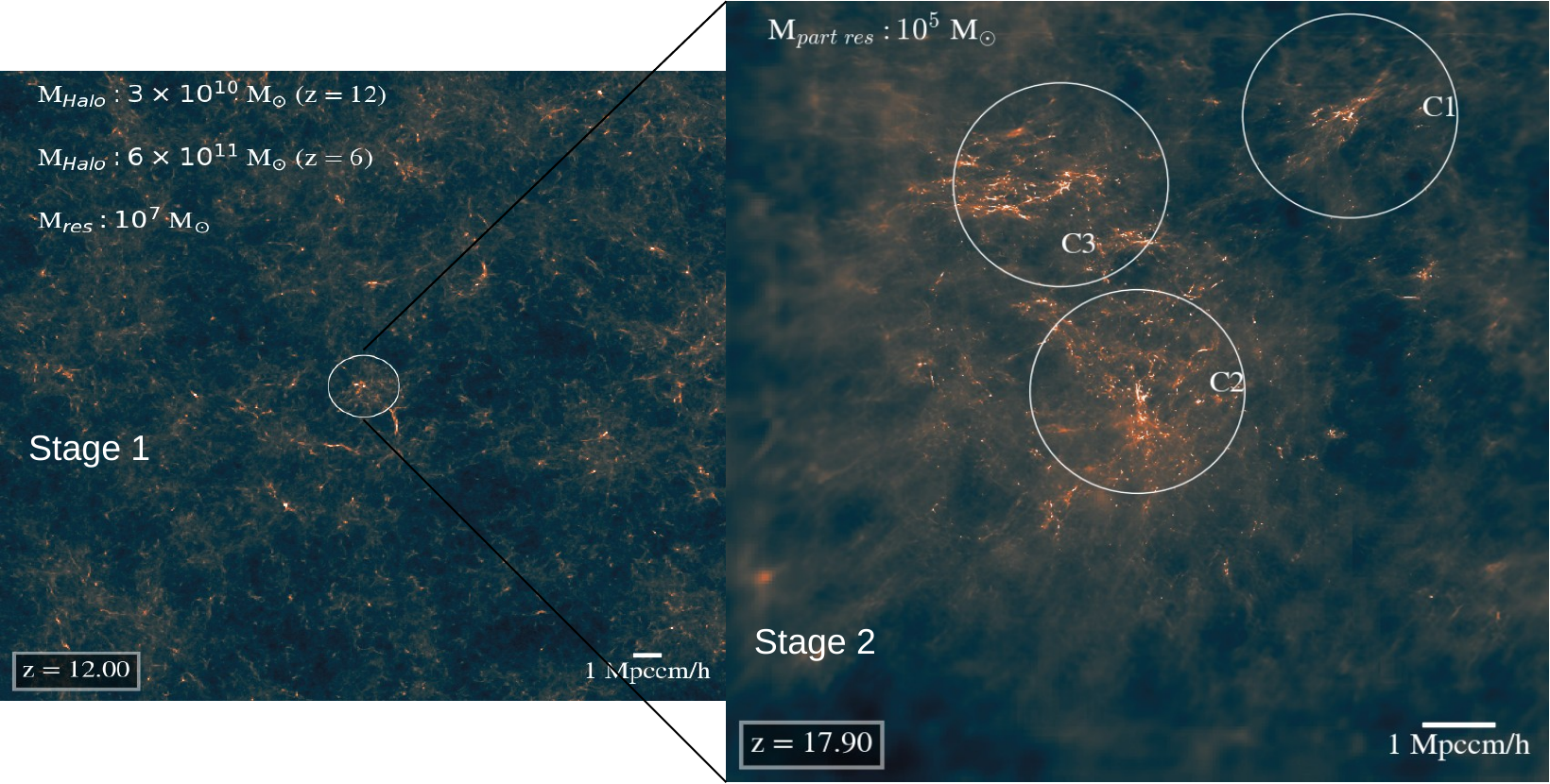}}
\caption{
  The initial setup required to identify the clusters needed to run our stage 3 simulations. In
  stage 1 (left hand side) a dark matter only simulation is evolved to a redshift of 6 in a box
  of side length 100 {\rm $h^{-1}$ Mpc}. Within this large box the most massive halo with the lowest tidal field
  is identified (white circle). The halo has a mass of M$_{halo} = 3 \times 10^{10}$ \msolar at z = 12
  and M$_{halo} = 6 \times 10^{11}$ \msolar at z = 6. Halos are identified and their mass determined
  using the \rockstar halo finder \citep{Rockstar}.
  For stage 2 we re-centre our initial conditions over this halo
  and rerun the simulations including baryonic physics, a 9 species chemical model and star formation
  in order to identify clusters of star formation in a $10^3$ \mpchv volume surrounding the halo found
  in stage 1. Three clusters are identified - as shown on the right hand side. For stage 3 we
  construct initial conditions surrounding each cluster with each cluster having a
  (comoving) volume of $1^3$ \mpchv. The clusters do not overlap as can be seen in the figure.
    The separation from the centre of one region to the centre of another varies from a minimum
    separation of 3.3 comoving Mpc/h to a maximum separation of 5.5 comoving Mpc/h.
  In total 31 haloes were observed to experience star formation
  during stage 2. The maximum mass
  resolution in Stage 1 is M$_{res} = 10^7$ \msolarc, while in Stage 2 the maximum mass resolution is
  M$_{res} = 10^5$ \msolarc. In stage 3 the maximum mass resolution is M$_{res} \sim 2.3 \times 10^3$
  \msolarc.
}
\label{Fig:InitialSetup}
\end{center} \end{minipage}
\end{figure*}

\section{Methods} \label{Sec:Methods}

\subsection{BlackDemon Simulation Suite} \label{Sec:BlackDemon}
\noindent The \blackdemon (Black Hole Demographics) simulation suite is a set of cosmological zoom-in simulations
targeting the progenitor haloes of a massive halo (M$_{halo} \sim 6 \times 10^{11}$ \msolar at z = 6)
in an overdense region of the early Universe (z $>$ 15)
(see Figure \ref{Fig:InitialSetup}). The \blackdemon simulations are run using the \enzo code \citep{Enzo_2014, Enzo_2019}
with a three stage setup required in order to achieve the mass resolution necessary to properly
resolve the progenitor haloes of massive star and black hole formation.\\
\indent \enzo has been extensively used to study the formation of structure in the early universe
\citep{Abel_2002, OShea_2005b, Turk_2012, Wise_2012b, Wise_2014, Wise_2019, Regan_2020}.
\enzo includes a ray tracing scheme to follow the propagation of radiation from
star formation and black hole formation \citep{WiseAbel_2011} as well as a detailed multi-species
chemistry model that tracks the formation and evolution of nine primordial species \citep{Anninos_1997,
  Abel_1997}. In particular the photo-dissociation of \molH is followed, which is a critical
ingredient for determining the formation of the first metal-free stars \citep{Abel_2000}. The simulation
  suite uses a Planck-like \citep{Planck_2014, Planck_2018} cosmology with the following
  parameters $\Omega_M = 0.2592, \Omega_{\Lambda} = 0.7408,
\Omega_b = 0.0487, H_0 = 67.7, \sigma_8 = 0.8159$ and n = 0.9667. \\

\subsection{Rationale}
\noindent The rationale guiding this simulation suite is to maximise resolution (both mass and spatial)
within a highly overdense region of the early Universe. The goal is to investigate both galaxy and
black hole formation in such a region. It is within these unusually overdense regions that
the seeds of MBHs may be expected to form more readily \citep{Regan_2020, Regan_2020b, Lupi_2021, Latif_2022, Chiaki_2023}.
While there is significant theoretical evidence to back this up observational evidence is still uncertain regarding
the high-z quasars at least \citep{Willott_2005, Kim_2009, Simpson_2014, Mazzucchelli_2017} but see \cite{Overzier_2022}
for tentative evidence for an overdensity near a high-z quasar.
The simulation suite consists of three stages. Stage one involved running a low resolution dark matter
only simulation to identify the Lagrangian region of a massive halo at high redshift. In stage two,
nested grids were used to gain higher mass resolution in the region surrounding the massive halo identified
in stage one. Individual haloes hosting early star formation were identified and then additional
nested grids were added for the final stage three simulations. For each of stage one, stage two and
  stage three the entire 100 $\rm{h^{-1}}$ Mpc box is evolved. What changes from one stage to another is the positioning and
  extent of the nested grids. For stage two the extent of the most refined nested grid is 10 $\rm{h^{-1}}$ Mpc, for
  stage three the most refined nested grid is 1 $\rm{h^{-1}}$ Mpc. All stages are initialised at z = 100.\\
\indent We therefore initially identified a massive
dark matter halo within a 100 \mpch volume, and by successively tracing back its progenitor haloes,
investigated star and black hole formation in that environment. The region surrounding
the formation of a massive halo is likely to experience far more mergers than a typical region and
as such is the ideal laboratory to investigate the potentially complementary effects of
both a rapid assembly history \citep{Yoshida_2003a, Fernandez_2014, Wise_2019} including
turbulent flows \citep{Latif_2022} and LW radiation \citep{Haiman_2006, Shang_2010} - all
of which have been advocated as mechanisms for forming SMSs and/or MBHs. 

\subsection{The Stage One and Stage Two Simulations}
As already noted the zoom-in regions were chosen as part of a three stage process.
Stage 1 involved running a large, low resolution, dark matter only simulation to find a massive
dark matter halo in the early
Universe. The initial conditions for this simulation were generated with MUSIC \citep{Hahn_2011}
within a 100 \mpch box run on a 2048$^3$ grid. This lead to a particle resolution of
M$_{part} \sim 8 \times 10^6$ h$^{-1}$ \msolar with a spatial resolution of approximately 5 \kpch (using
4 levels of adaptive refinement). The simulation was evolved to a redshift of z = 6. At this point
the \rockstar halo finder \citep{Rockstar} was used to locate massive haloes within the box. Rather
than simply choosing the most massive halo in the box we choose instead the halo with the lowest
tidal field \citep[e.g.][]{DiMatteo_2017} among the ten most massive haloes. Our selected halo had a
mass of M$_{halo} \sim 6.1 \times 10^{11}$ \msolar at z = 6 and a mass of M$_{halo} \sim 3 \times 10^{10}$
\msolar at z = 12. This halo corresponds to a greater than 4 $\sigma$ fluctuation
in the primordial density field. \\
\indent Having found the massive halo we used this to centre our next simulation and placed
six static nested grids around the halo. The finest grid was 10 \mpch on the side and gave
a maximum particle resolution of M$_{part} \sim 10^5$ h$^{-1}$ \msolarc. This stage 2 simulation was
run with full baryonic physics turned on including the star formation prescription described in
\S \ref{Sec:StarFormation}. No feedback was employed
in this stage however. The goal of this stage was simply to identify the clustered regions inside this
sub-volume within which (massive) star formation initially takes place. At this particle mass resolution
only haloes with masses of a few times $10^6$ \msolar could be resolved and so while this simulation cannot
determine the true pathways to early star formation it nonetheless provided us with the information
we need to continue to stage 3. The stage 2 simulations were evolved to z = 17.9.\footnote{We terminated the simulations at this
point as our initial goal is to run the Stage 3 simulations to z = 18.} \\
\indent At z = 17.9 we had 31 haloes undergoing active star formation inside this 10 \mpch sub-region
($\sim 0.5$ \mpch physical). This allowed us to identify three distinct ``clusters'' of
star formation (see Figure \ref{Fig:InitialSetup}). These three clusters provided us with the
initial conditions required to begin our stage 3 simulations. For stage 3
an additional two static nested grids (so eight in total) were placed around each cluster. This gives
three different initial conditions each centred on a different cluster (named Cluster1, Cluster2 \& Cluster3).
Additionally, we created initial conditions for a 1 \mpch ``ControlRegion'' which is not connected to the overdense region
in any way and was instead selected from a region of cosmic mean density chosen randomly from the
parent box. The ``ControlRegion'' is run under a setup identical to the cluster regions and acts as
a control to compare against the results from the overdense region.
Within the finest static mesh encompassing each cluster (including the ``ControlRegion'')
the maximum particle resolution is now M$_{part} \sim 2.3 \times 10^3$ \msolarc. Each region
has a side length of 1 \rm $h^{-1}$ Mpc \footnote{We initially tried larger high resolution volumes but
they proved intractable}. Within this region adaptive mesh refinement is permitted
down to a maximum depth of 20. This corresponds to a maximal spatial resolution of
$\Delta x \sim 0.05$ pc (physical, 1.1 pc (comoving)) at z = 20. 

\subsection{Stage 3 simulations}
\noindent The stage 3 simulations, which are currently still running, are designed to be able to
resolve all of the pathways to MBH formation as self-consistently as possible.  The
refinement criteria used here are based on three physical measurements: (1) The dark
matter particle over-density, (2) the baryon over-density and (3) the Jeans length. The first two
criteria introduce additional meshes when the over-density
(${\Delta \rho \over \rho_{\rm{mean}}}$) of a grid cell with respect to the mean density exceeds 8.0
for baryons and/or DM. Furthermore, we set the \emph{MinimumMassForRefinementExponent} parameter
to $-0.1$ making the simulation super-Lagrangian and therefore reducing the threshold for
refinement as higher densities are reached. For the final criteria we set the number of cells
per Jeans length to be 4 in these runs. 

The stage 3 simulations require our full physics model implementation to accurately
track massive star and black hole formation as well as accretion and feedback. 
Our particle resolution of M$_{part} \sim 2.3 \times 10^3$ \msolar allows us to identify all
haloes with masses above approximately $10^5$ \msolarc. The minimal Jeans mass for gravitational
collapse is expected to be above this threshold 
mass \citep{Bryan_1998, Fuller_2000, Machacek_2001, Yoshida_2003a, Skinner_2020, Schauer_2021, Kulkarni_2021}
meaning that these simulations are capable of
resolving the first mini-haloes. The focus on an overdense region means that these
clusters and the individual haloes within them are very likely to be subject to dynamical heating
from repeated minor mergers \citep{Yoshida_2003a, Fernandez_2014, Wise_2019},
turbulent flows \citep{Latif_2022} as well as merger triggered star formation \citep{Hopkins_2013}
all of which have been shown to drive massive star formation in
haloes which are already above the threshold for gravitational
collapse. The stage 3 simulation suite include a full multi-species model including
a primordial chemical network of 9 species with the detailed chemical network solved using the \grackle library
\citep{Grackle}. \grackle models self-shielding of \molH using the prescription given by \cite{WolcottGreen_2011},
shielding of \molH due to intervening neutral Hydrogen is currently not part of \grackle and could be
added in the future to improve the accuracy of modelling \molH abundances \citep[e.g.][]{Schauer_2015}.
Star formation is tracked self-consistently and is detailed below. 

 \subsection{Subgrid Star Formation \& Black Hole Prescription} \label{Sec:StarFormation}
 \noindent In order to resolve star formation in the collapsing target haloes we set the maximum
 refinement level of the stage 3 simulations to 20 which corresponds to a maximum spatial resolution
 of $\Delta x \sim 0.05$ pc at z = 20.  \\
 \indent To model star formation within the collapsing gas cloud we employed a star
 formation criteria using the methodology first described in \cite{Krumholz_2004} and more
 recently in \cite{Regan_2018a, Regan_2018b, Regan_2020b}. 
 Stars are formed when all of the following conditions are met:
\begin{enumerate}
\item The cell is at the highest refinement level
\item The cell exceeds the Jeans density 
\item The flow around the cell is converging 
\item The cooling time of the cell is less than the freefall time
\item The cell is at a local minimum of the gravitational potential
\end{enumerate}
The simulation results reported here are
at the point where initial gravitational collapse has been attained in all four regions and
star formation and evolution is subsequently tracked in the ControlRegion, Cluster2 and Cluster3.
While we track the initial evolution of four separate regions we only track the onset of star and black hole formation in
the ControlRegion, Cluster2 and Cluster3 due to the computational expense of the simulations. 
Within the three regions in which we track star formation 
three of the first five stars to have formed have collapsed into MBHs. \\
\indent In tracking star formation in such an overdense environment the question becomes what type
of star is formed at the outset of star formation and
how does that star evolve over time? The type of star that forms is based on the following logical
sequence:
\begin{enumerate}
\item If the (gas-phase) metallicity\footnote{Note that all metallicities
referred to in this paper are gas phase metallicities.} is greater than $10^{-4}$ \zsolar then a
  cluster of PopII stars is formed (represented by a single particle)
\item If the metallicity is less than  $10^{-4}$ \zsolar and the accretion rate onto the cell is
  greater than 0.01 \msolaryr then a SMS forms (note that for initial formation we require an accretion
  rate a factor of ten above the critical accretion rate\footnote{we take the critical accretion rate for SMSs to be 0.001 \msolaryr \citep{Haemmerle_2018}}
  to ensure we are well inside the accretion rates necessary
  for SMS formation)
\item If the metallicity is less than $10^{-4}$ \zsolar and the accretion rate onto the cell is
  less than 0.01 \msolaryr and the \molH fraction is greater than $5 \times 10^{-4}$ then a PopIII star
  forms
\item Otherwise no star particle is created and the code continues to check for star formation regions
\end{enumerate}
  
\begin{table*}[!t]
  \centering
  \begin{tabular}{ |p{3cm}|p{2cm}|p{3cm}|p{4cm}|p{4.1cm}| } 
    \hline
    \textbf{Cluster Name} & \textbf{Current \newline Redshift} & \textbf{First Star \newline Formation Redshift} & \textbf{\# Haloes \newline (M$_{Halo} \ge 1 \times 10^5$ \msolarc)} & \textbf{Maximum Halo Mass [\msolarc]} \\
    \hline
    Cluster1 & 22.28 & - & 1361 & 2.65 $\times 10^6$ \\
    \hline
    Cluster2 & 22.80 & 23.06 & 2826 & 1.72 $\times 10^7$ \\
    \hline
    Cluster3 & 22.10 & 22.31 & 3522 & 2.36 $\times 10^7$ \\
    \hline
    ControlRegion & 19.8 & 20.00 & 321 & $1.81 \times 10^6$ \\
    \hline
  \end{tabular}
  \caption{\noindent Cluster statistics at the current time of reporting. From left to right we have the cluster name, the current cluster redshift
    at the time of reporting, the redshift of first star formation in each cluster (note we do not follow Cluster1 to the point of star formation)
    the number of resolved haloes within the cluster region and the maximum halo mass in the region
    at the cluster redshift.} \label{Table:HaloStats}
\end{table*}

\begin{figure*}
\centerline{
  \includegraphics[width=9.0cm]{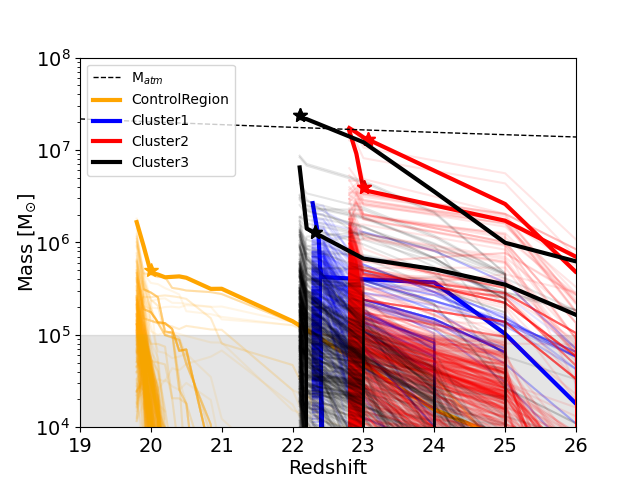}
  \includegraphics[width=9.0cm]{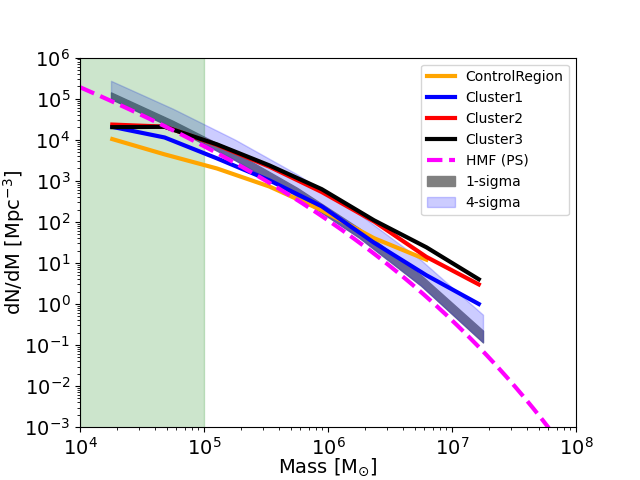}}

\caption{
  \textit{Left Panel:} The growth of a large selection of haloes from each of Cluster1, Cluster2, Cluster3 and the
  ControlRegion. Each cluster is identified by colour - blue for Cluster1, red for Cluster2,
  green for Cluster3 and orange for the ControlRegion. The star forming haloes (or the first halo
  to undergo gravitational collapse in the case of Cluster1)
  are marked by the thick line and star markers indicating the redshift of first star formation.
  At the redshifts to which the
  simulations have now evolved most resolved haloes are between $10^5$ \msolar and $10^7$ \msolarc.
  \textit{Right Panel:} The halo mass function from each of Cluster1, Cluster2, Cluster3 and the
  ControlRegion at the final output in each case. Overlaid on top (magenta dashed line) is the
  halo mass function from the publicly available HMF library \citep{Murray_2013} as well as the 1-sigma
  and 4-sigma bands from the HMF computed halo mass function.
}
\label{Fig:MassRedshift}
\end{figure*}

The initial masses assigned to stars are determined by their type. SMSs accrete material and hence their
masses are calculated self-consistently. PopII and PopIII stars have initial masses assigned at birth.
All PopII (clusters) are assigned masses based on the overdensity in the cell, compared to the Jeans
density. The minimum PopII mass is set to be $10^3$ \msolarc. Pop II star particles represent a star cluster of some total mass
  and an assumed normal (Gaussian) IMF. To generate the PopII star particles a sphere is created around the maximum density cell
  and a fraction c$_* = 0.07 \times \rm{f}_{\rm{cold}}$ of the cold gas, $\rm{f_{cold}}$, is converted into stars
  (with a minimum cluster mass of 1000  \msolarc).PopIII stars on the otherhand have initial
masses assigned according to the same process as used in the \renaissance suite of simulations
\citep[e.g][]{Chen_2014}. In this case masses are chosen randomly from an initial mass function with a
functional form given by
\begin{equation}
  f(log M) dM = M^{-1.3} \rm{exp} \big [ - \big (\frac{M_{char}}{M} \big) \big ] dM
\end{equation}
This form behaves as a power-law at $\rm{M > M_{char}}$ and is exponentially cutoff below that mass \citep{Chabrier_2003}.
We choose $M_{char} = 40$ \msolar consistent with recent result of PopIII formation simulations in the literature
\citep{Turk_2009, Clark_2011a, Stacy_2014, Hirano_2014}. The PopIII mass range runs from 1 \msolar up to 300 \msolarc.
The star formation criteria envoked here combined with our maximum spatial resolution results in a typical Jeans Density
  at the centre of the collapsing structures of a few times $10^6 \ \rm{cm}^{-3}$. It also needs to be emphasised that this
  density resolution is insufficient to probe the true spatial scale of star formation. In practice our results should therefore
  be interpreted in this light in the knowledge that subsequent fragmentation may reduce the masses of the individual stars. Although
  there is good reason to believe that the regions modelled here, which experience high gas inflows, form very massive central objects even
in the face of fragmentation \citep[e.g.][]{Reinoso_2023}.
Black holes, when formed, form with the same mass as the progenitor star. \\
\indent  Subsequent to a star forming and being assigned a type, accretion onto the star is tracked
if the star is either a SMS or a black hole. We do not have sufficient resolution in these
simulations to track accretion onto PopIII stars (which are much more compact than SMSs). PopII
clusters also do not accrete. As discussed in the Introduction SMSs are stars
with low surface temperatures that are appropriate for main sequence SMSs and less massive proto-stars on the
Hayashi track. Rapidly accreting (i.e. $\dot{M_*} \gtrsim 0.001$ \msolarc/yr) protostars
  carry large amounts of entropy (hot accretion) into the stellar interior. The stellar radius
  monotonically increases as the stellar mass increases obeying an analytic mass-radius relation
  \citep{Hosokawa_2010, Hosokawa_2012, Hosokawa_2013}.

  \begin{equation}
    R_* \approx 2.6  \times 10^3 R_{\odot} \Big( \frac{M_*}{100 M_{\odot}}\Big){1/2}
  \end{equation}
  where $R_*$ is the stellar radius and $M_*$ is the stellar mass. Note that the relation
  is independent of the actual accretion rate. The stellar interior remains inhomogeneous and
  subsequently  contracts radiating energy away. A surface layer containing a small fraction of
  the mass inflates leading to a puffy SMS with low effective temperatures. The expansion continues
  until the radius eventually begins to contract when the mass of the star exceeds
  $M_* \gtrsim 3 \times 10^4$ \msolarc. This occurs because H$^-$ bound-free opacity, which keeps
  the stellar surface temperature locked at close to 5000 K becomes unavailable as the density in
  the surface layer drops below $10^{-11}\ \rm{g\ cm^{-3}}$. Nonetheless,
  the radius at this stage of its evolution is approximately 100 AU. While this stellar radius is
  still below our resolution scale we nonetheless calculate the accretion rate flowing radially onto
  a region with a radius of 4 cell widths around the particle and use the accretion formalism
  of \cite{Krumholz_2004} to approximate the accretion rate onto the stellar surface for SMSs. \\
  \indent Assuming a star is designated as a SMS initially, as the simulation evolves the
  accretion rate onto a SMS can drop below the critical rate
  in which case the star contracts to the main sequence and becomes a PopIII star. This reflects
  the contraction of the Pop III star to the main sequence on the Kelvin-Helmholtz timescale.
  For the simulation resolution of \blackdemon we do not have sufficient resolution to track accretion onto PopIII stars
  and hence once the star becomes a PopIII star (either at formation or as a result of transitioning
  from a SMS to a PopIII star) it will remain a PopIII star until the end of its stellar lifetime.
  Pop III stars are modelled assuming a blackbody spectrum with an effective temperature of
  T$_{eff} = 10^5 $ K \citep{Schaerer_2002} (no mass loss rates)
  while SMSs are modelled by assuming a blackbody spectrum with an effective temperature
  of T$_{eff}$ = 5500 K \citep{Hosokawa_2013}. The luminosity rates for the PopIII star as given
  by  \cite{Schaerer_2002} have recently been verified by \cite{Haemmerle_2017b} who recover the
  rates of \cite{Schaerer_2002} for cases where the accretion rate is below the critical rate.\\
  \indent At the end of its stellar lifetime (for both SMSs and PopIII stars) the star
  transitions into a black hole particle and depending on its mass perhaps through the
  intermediate stage of a supernova explosion. \\
  \indent SMS always collapse directly to a black hole \citep{Woods_2017} with the
  black hole retaining the same mass as the final SMS mass. PopIII stars in the
  range $20$ \msolar $< \rm{M_{PopIII}} < 40.1$ \msolar undergo a type II supernova explosion and
  subsequently transition into a black hole. PopIII stars in the
  range $140$ \msolar $< \rm{M_{PopIII}} < 260$ \msolar undergo a pair instability explosion with
  no remnant and the particle is removed from the simulation. PopIII stars in the
  range $40.1$ \msolar $< \rm{M_{PopIII}} < 140$ \msolar and those with masses in excess of 260 \msolar
  directly collapse into black holes of the same mass. \\

  \subsection{Subgrid Feedback} \label{Sec:Feedback}
  \noindent The radiative feedback from the stellar population depends on the stellar type.
  We use mass-dependent infrared, Hydrogen ionizing, LW and Helium ionising luminosities and lifetimes of both
  the PopII and Pop III stars from \cite{Schaerer_2002}. For PopIII stars the radiative feedback, which is propagated
  through the simulation using the in-built ray tracing formulation of \cite{WiseAbel_2011},
  models the radiation in five radiation bins from the infrared, to LW, to Hydrogen ionising
  radiation to both singly Helium ionising and doubly Helium ionising radiation. The energies assigned to each radiation
  bin are taken from \cite{Schaerer_2002}. In a similar vein the radiative feedback from
  metal enriched PopII stars is also taken from  \cite{Schaerer_2002} with radiation in this
  case spread across four radiation bins from LW, to Hydrogen ionising, to single and
  doubly ionising Helium radiation. 
  SMSs are modelled by assuming a blackbody spectrum with an effective temperature of T$_{eff}$ = 5500 K \citep{Hosokawa_2013}.
  The radiation spectrum for a SMS therefore peaks in the infrared as opposed to the UV for Pop III stars.
  For the specific luminosity of the SMS we take a characteristic mass of 500 \msolar and apply
  the contribution from the non-ionising photons only \citep{Schaerer_2002}. The SMS luminosity
  changes as mass is accreted and the total luminosity then
  scales up as the mass increases. When SMSs stars transition into PopIII stars we continue to model
  the radiative feedback below the Hydrogen ionisation edge using ray tracing. To limit the computational
  challenges we do not model the ionising radiation. Instead the feedback is approximated using a thermal feedback model where the
  gas surrounding the star is heated to $T = 10^5$ K hence ionising all of the in-situ gas. This approximation is likely to be robust given
  that the ionising radiation is not expected to break out from the immediate vicinity of the star given the large column densities
  \citep[e.g.][]{Woods_2021, Jaura_2022}.\\
  \indent To model the radiative feedback from black hole particles
  we assume a multi-colour disk for the accretion disk and then a fit a corona with a
  power law \citep[e.g.][]{Done_2012}. We divide the energy radiated equally between the multi-colour
  disk and the power law component.
  For black holes we split the radiation into five
  energy bins from infrared up to hard X-rays. The energy bins used are 2.0 eV, 12.8 eV, 19.1eV,
  217.3 eV and 5190 eV with the actual value of the luminosity at each timestep determined by the
  accretion rate at that timestep - if there is no accretion then there is no feedback. The fractional energy in each
  energy bin is then determined by the accretion rate onto the black hole and the mass of the black hole.
  We do not model kinetic or mechanical feedback
  from black holes in this suite of simulations. \\

\begin{figure}
\centering
\centerline{
  \includegraphics[width=10.0cm, height=8cm]{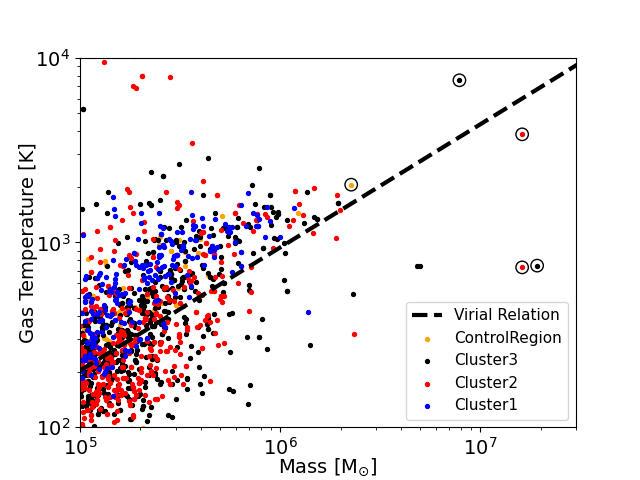}
}
\caption{
  Halo mass as a function of gas temperature at the virial radius of each halo. Each cluster
  is coloured uniquely as described in the legend. Those haloes which have collapsed and
  are beginning to form stars are circled to allow for easy identification. The haloes, in general,
  follow a trend of increasing temperature with mass in this regime. The five star forming haloes
  from the ControlRegion, Cluster2 and Cluster3 are marked by the open circles.
  The star forming massive haloes have temperatures that are impacted by (stellar) feedback
  in some cases. We also show Eqn \ref{Eqn:Tvir}, as a dashed line, showing that the haloes
  follow very closely the expected virial relation i.e. $\rm{T_{vir} \propto M_{halo}^{2/3}}$. We calculate the virial relation at z = 20.
}
\label{Fig:MT}
\end{figure}

\section{Results} \label{Sec:Results}

\noindent This first paper in the \blackdemon (Black Hole Demographics) series examines the formation of dark matter haloes and
embryonic galaxies in an overdense region. This analysis investigates early halo formation
immediately prior to gravitational collapse which leads to the first star in each region and the
subsequent evolution for more than two million years\footnote{Beyond the formation for the first
star in each clustered region we follow only the evolution of Cluster2 and Cluster3. This is because
Cluster2 and Cluster3 show the highest abundance of rapidly growing haloes which is of most interest
in this study. Additionally we follow the evolution of the ControlRegion for comparison.}.
Galaxies in this overdense environment are
predicted to be influenced by the rapid assembly of dark matter haloes in such an environment and
as such may not necessarily follow, in all cases, the predictions from theory when the \molH fraction
is taken as the primary driver of catastrophic cooling and collapse. We begin by reviewing the
relevant physical processes that are predicted to drive cooling in this regime and then compare these
results to the results of our numerical simulations across each of our four regions (Cluster1, Cluster2,
Cluster3 and the ControlRegion). \\
\indent Table \ref{Table:HaloStats} outlines the current conditions in each region (i.e. at the point of
gravitational collapse of the first halo in Cluster1 and approximately 2 Myr after the formation of the first star
in each of the ControlRegion, Cluster2 and Cluster3). The table gives the number of resolved (at the current redshift)
haloes in each cluster, the cluster redshift at the time of reporting, the redshift of first star formation
and the maximum halo mass at this epoch in each case.
The left panel of Figure \ref{Fig:MassRedshift} shows the mass of a selection of haloes against
redshift\footnote{subhaloes are largely excluded leaving only haloes with unique assembly histories}.
Each Cluster is coloured uniquely (blue for Cluster1, red for Cluster2, black for
Cluster3 and orange for the ControlRegion). The star forming haloes (or the first to
undergo gravitational collapse in the case of Cluster1) are identified by thick
lines, with star markers indicating the redshift of first star formation, in each case.
What is immediately clear is that the regions in the overdense regions all have higher collapse
masses compared to the ControlRegion. The most massive halo currently exists in Cluster3 and has a
mass approximately an order of magnitude larger than than that in the ControlRegion.
The values below M$_{Halo} = 10^5$ \msolar are shaded as haloes below this scale are not well
resolved in our simulations (M$_{part} \sim 2.3 \times 10^3$). Nonetheless, we show their growth
history through this mass range for completeness. \\
\indent To compliment the mass growth plot in the right
hand panel of Figure  \ref{Fig:MassRedshift} we plot the halo mass function for each cluster. We
additionally plot the halo mass function for a press-schecter mass function \citep{PressSchecter_1974}
using the HMF library \citep{Murray_2013} (magenta dashed line) as well as the 1-sigma and 4-sigma
bands. Both Cluster2 and Cluster3 show significant deviations from the PS mass function
reflective of the fact that they are rare manifestations of mean structure formation. The
ControlRegion and Cluster1 align more closely with the PS mass function. Note that all four regions
show deviations from the analytical fits at low ($\lesssim 2 \times 10^5$ \msolarc) and high
($\gtrsim 5 \times 10^6$ \msolarc) halo masses but this is due to poor statistics at these extremes.
We will see again in \S \ref{Sec:OverdenseRegion} that Cluster2 and Cluster3 exhibit strong
characteristics of evolving inside a highly overdense region.

\subsection{Analytical Predictions}
\noindent The first mini-haloes in our Universe are expected to form with masses greater than $10^6$ \msolar
\citep{Bryan_1998} with the cooling within these first small galaxies driven by the presence of \molH \citep{Tegmark_1997}.
It is therefore instructive to consider how the predicted \molH cooling timescale, $\tau_{H_2}$, evolves. The timescale for
\molH cooling can be written as:

\begin{equation} \label{Eqn:H2CoolingTime}
  \tau_{H_2} = \frac{1}{\gamma - 1} \frac{\rm{n}_{\rm{gas}}}{\rm{n_H^2} f_{H_2}} \frac{\rm{k_{B}} \rm{T_{vir}} }{\Lambda_{H_2} }
\end{equation}

where $\gamma$ is the adiabatic constant, $\rm{n_{gas}}$ is the gas number density, $\rm{k_{B}}$ is the Boltzmann constant,
$\rm{T_{vir}}$ is the halo virial temperature, $\rm{n_H}$ is the hydrogen
number density, $f_{H_2}$ is the \molH fraction and $\Lambda_{H_2} (T) $ is the \molH cooling function given by
$\Lambda_{H_2} (T) = 10^{-27.6} (T_{vir} / 100)^{3.4}$ erg s$^{-1}$ cm$^3$. This fitting is valid for temperatures in the
range $120 \ \rm{K} \le T \le 6400 \ \rm{K} $ \citep{Trenti_2009}. The virial temperature of a halo can be calculated analytically
as

\begin{equation} \label{Eqn:Tvir}
  \rm{T_{vir} = \kappa \Big(\frac{\mu}{0.6} \Big) \Big(\frac{\Omega_m}{\Omega_{mz}} \frac{\Delta_{vir}}{18 \pi^2} \Big)^{1/3} \Big(\frac{h \ M_{vir}}{10^8 M_{\odot}} \Big)^{2/3}
  \Big(\frac{1+z}{10} \Big)}
\end{equation}
where $\kappa =  1.98 \times 10^4$, $\mu$ is the mean molecular weight, $\Omega_m$ is the cosmological matter density,
$\Omega_{mz} = {{\Omega_m (1 + z)^3} \over {\Omega_m (1 + z)^3 + \Omega_{\Lambda}}}$, $\Delta_{vir} = 18 \pi^2 + 82d - 39 d^2, d = \Omega_{mz} - 1$ and $ M_{vir}$ is the halo
virial mass \citep{Barkana_2001, Lupi_2021}.

For each dark matter halo identified in our Stage 3 simulations (halos are identified using the \rockstar halo finder) we can calculate the
\molH cooling time for each halo since we have the required information (i.e. \molH fraction, halo virial mass and virial temperature). Since our
simulations have a full hydro component we also know the actual cooling time of the halo as calculated using \texttt{Grackle}. We are therefore
able to use the analytical predictions to directly compare against the full hydrodynamical results in order to observe the impact of local
environmental conditions on gravitational collapse inside the first structures and compare
them against analytic predictions.\\
\indent As the hierarchical structure within a cosmological region develops, halos grow through a series of mergers. The frequency and magnitude of the merging
process effects the ability of the gas to cool as kinetic energy is added to the gas via the dark matter build up. The energy injected as a result of this
process can be referred to as ``dynamical heating'' and was initially investigated numerically by \cite{Yoshida_2003a}. In their N-body/SPH simulations they found that
haloes that exceeded a critical growth rate generated sufficient heating within a halo to offset \molH cooling and hence delay gravitational collapse. The
critical growth rate is given by

\begin{multline} \label{Eqn:crit}
  \left.\frac{dM}{dt}\right\vert_{crit} \approx \Big( \frac{n_H}{10^2}\Big)  \Big( \frac{M}{10^6 M_{\odot}} \Big)^{1/3} \Lambda_{H_2} \Big( \frac{T}{1000 \ \rm{K}} \Big) \\ \Big( \frac{f_{H_2}}{10^{-4}} \Big)
  \Big (\frac{1}{\alpha_{vir} (T/1000 \ \rm{K}, M/10^6 \ \rm{M_{\odot}})} \Big) \Big( \frac{\gamma - 1}{k_B} \Big) \ \ \rm{M_{\odot}/yr}
\end{multline}

where $\alpha_{vir} = T_{vir} /(M_{vir}/M_{\odot})^{2/3}$ and n$_{H}$ is the number density of Hydrogen at the virial radius. Using the canonical values above the critical halo growth rate is
approximately

\begin{equation}  \label{Eqn:CriticalRate}
\frac{dM}{dt}\vert_{crit} \approx \ 1 \ \rm{M_{\odot}/yr}
\end{equation}

\subsection{Halo Characteristics within the Overdense Regions} \label{Sec:OverdenseRegion}
\begin{figure}
  \centerline{
    \includegraphics[width=9cm]{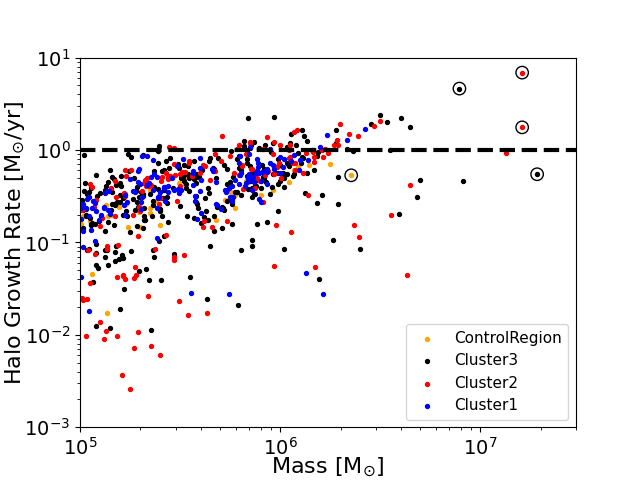}
  }
\caption[]{\label{Fig:dMdt_CanCool}
   The growth rate of each halo plotted against its mass. Each cluster is
   distinguished by its colour as described in the legend.
   Only those haloes for which the calculated
   cooling time is less than a tenth of the Hubble time are shown. The growth rate shown
   is the instantaneous growth rate (i.e. the growth rate of each halo at this snapshot time).
   The star forming haloes (circled) in Cluster2 and Cluster3 all have relatively high growth rates.
   The star forming halo in the ControlRegion has a more modest growth rate but its halo mass is significantly
   lower and the halo resides in a more 'normal' environment. 
}
\end{figure}

\noindent We begin our quantitative analysis of the haloes by examining the halo mass - temperature plane of
the haloes from our simulation clusters. In Figure \ref{Fig:MT} we show the halo mass against the
halo temperature, haloes from different clusters are coloured as indicated in the legend (orange for
the ControlRegion, blue for Cluster1, red for Cluster2 and
black for Cluster3). The gas temperature is calculated by taking the density weighted average value
at the halo virial radius.
From Figure \ref{Fig:MT} we can see that the halos follow a trend of increasing temperature with
mass as expected in this low mass (density) regime. The star forming haloes from the ControlRegion,
Cluster2 and Cluster3 are marked by open circles and have virial temperatures of between a few
hundred Kelvin up to almost $10^4$ Kelvin. The higher temperatures seen in some of these massive,
star forming, haloes is due to the impact of stellar feedback (see \S \ref{Sec:Onset}). \\
\indent In Figure \ref{Fig:dMdt_CanCool} we plot the halo growth rate as a function of halo mass.
Furthermore, we include only those haloes which have calculated cooling times
(from Grackle) which are a factor of 10 shorter than the Hubble time at this redshift. We choose
cooling times which are only a tenth of the Hubble time given the very early age of the
Universe at this epoch and hence it's comparatively short age. Comparing to the full Hubble time
at this epoch would be misleading and we use the, admittedly somewhat arbitrary,
value of one tenth the Hubble value to be conservative. \\
\indent The overall distribution shows a large scatter with values ranging from $10^{-3}$ \msolaryr to
greater than 1 \msolaryrc. The dashed horizontal line shows the critical growth rate that must
be achieved to drive dynamical heating which can offset cooling due to \molHc. The critical growth
rate is for a halo of mass $10^6$ \msolarc, temperature,
T = 1000 K, and a \molH fraction of $f_{H_2} = 10^{-4}$. A handful of haloes
(predominantly from Cluster2 and Cluster3) show mass growth rates exceeding this critical value.
Additionally, three of the five star forming haloes have growth rates significantly above the
critical rates while also being in a regime where atomic cooling may not yet be fully active given
the haloes relatively low temperatures. The halo growth rates in this case (i.e. dM/dt) is computed
by calculating the finite difference in halo masses between the current and
previous data output. \\

\subsection{Determining the impact of dynamical heating}
\begin{figure}
\centering
\centerline{
  \includegraphics[width=9.0cm]{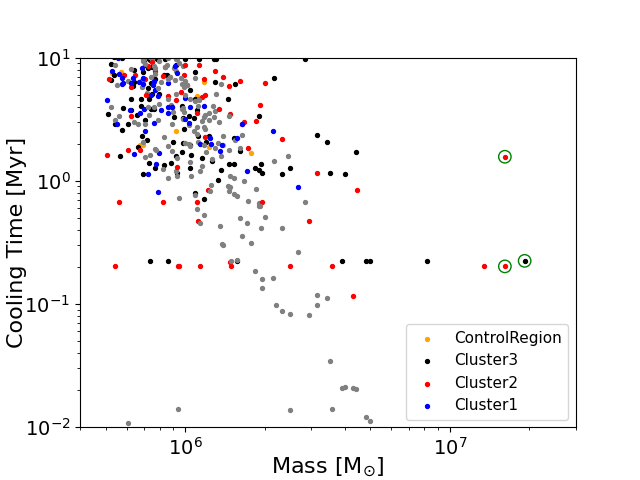}
}
\caption{
  The cooling times from the analytic \molH cooling function
  (Equation \ref{Eqn:H2CoolingTime}, grey dots)
  and from the numerically calculated values inside each halo (coloured dots).
  The analytically predicted \molH cooling times for some halos, particularly those at higher masses,
  are significantly below those
  actually calculated numerically. Rather than cooling, the haloes are being dynamically heated
  due to rapid assembly. The star forming halos from Cluster2 and one from Cluster3 are circled. Cooling
  times in these haloes are additionally impacted by (radiative) feedback. In particular the
  cooling times for the star forming halo in the ControlRegion and the additional star forming halo from Cluster3
  are impacted by feedback and are of order a hundred Myr and extend off the plot. 
}
\label{Fig:H2CoolingVActual}
\end{figure}

\noindent We now attempt to extract the effect of dynamical heating in such an overdense region. We
begin by examining the predicted cooling time due to \molH cooling on each halo and
compare that against the actual cooling time for each halo. When retrieving the \molH
  fractions and the (grackle calculated) cooling times from the simulation data all quantities
  are extracted from the simulation within a 10 pc sphere centred on the densest point in the
  halo. Both the \molH and cooling times are taken as density weighted averages.
The timescale for \molH cooling is given
by Eqn \ref{Eqn:H2CoolingTime}. In Figure \ref{Fig:H2CoolingVActual} we plot the
predicted cooling time for each halo in grey. Over-plotted onto this we plot (colouring each region
again) the calculated cooling time for each halo. Again we only plot the cooling
times for haloes with cooling times shorter than one tenth of the Hubble time. There is substantial
scatter at halo masses
below  M$_{Halo} = 5 \times 10^5$ \msolar and hence we only show haloes above this mass threshold.\\
\indent Above M$_{Halo} = 10^6$ \msolar we can clearly discern the impact of rapid assembly.
The grey points, which represent the analytical predictions for each halo based on the halo \molH
fraction systematically under-predict the gas cooling times (particularly for the
rapidly growing haloes in Cluster2 and Cluster3). If the actual cooling times followed a similar
trend to the predicted cooling times then we would see no coloured points towards the right hand
side of the plot. However, a number of coloured points (from Cluster2 and Cluster3) are seen to extend
to the right hand side. This is because these haloes are not cooling as predicted and have
longer than expected cooling times. The actual calculated values can be up to several Myr longer
and given the halo growth rates this can lead to much more massive haloes
developing before the actual cooling times are reached.\\
\indent The increase in the cooling time of the gas means that the halo can continue to grow
without collapsing. This has been shown in previous works, which in many cases looked
at specific processes which increased cooling times, sometimes somewhat artificially, leading to the
formation of
(super-)massive stars through both the dissociation of the \molH fraction
\citep[e.g.][]{Latif_2014b, Latif_2016a, Regan_2016a, Regan_2017}, through the
impact of baryonic streaming velocities \citep{Tanaka_2014, Latif_2014c, Hirano_2017, Schauer_2017}
and through rapid assembly \citep{Fernandez_2014, Wise_2019, Regan_2020b}.

\begin{figure}
\centering
\centerline{
  \includegraphics[width=9.0cm]{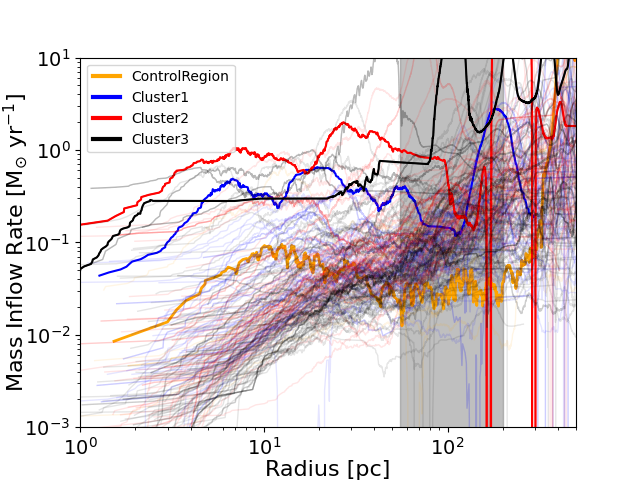}
}
\caption{The mass inflow rates for each of the haloes identified as having cooling times
  shorter than one tenth of the Hubble time. The inflow rates cover a large range as expected
  given the diversity of environmental factors. The grey shaded region denotes the span
  in virial radii for the haloes plotted. Within this shaded regions the average mass inflow
  rate is approximately 0.2 \msolaryrc.}
\label{Fig:HaloAccretionRates}
\end{figure}

\begin{figure}
\centering
\centerline{
  \includegraphics[width=9.0cm]{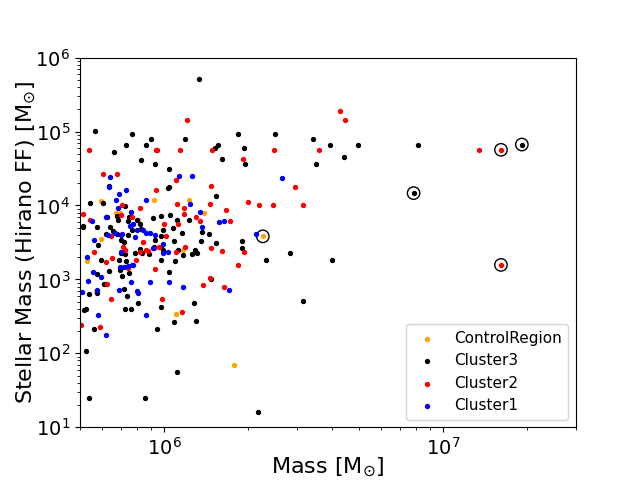}
}
\caption{The fitting function from \cite{Hirano_2014}
  ($ \rm{M_{popIII}} = 100 \rm{M_{\odot} \frac{\dot{M}_{virial}}{1.2 \times 10^{-3} M_{\odot} yr^{-1}}}$)
  predict stellar masses in the
  range $\rm{M_{*} \sim 100}$ \msolar up to $\rm{M_* \gtrsim 10^5}$ \msolar consistent
  with what would be predicted for haloes with high accretion rates. These numbers should be
  viewed as upper limits as both fragmentation and limitations on stellar growth are difficult to
  quantify. Nonetheless, it appears we are tracing the tail of the PopIII IMF. The circled haloes again
  identify the five haloes currently undergoing star formation. Based on their mass infall rates
  (calculated at the virial radius) the fitting function predicts stellar masses between
  approximately $10^3$ \msolar and $10^5$ \msolarc. This is broadly consistent with, if
  somewhat higher than, what our detailed numerical simulations find in practice.}
\label{Fig:HiranoRates}
\end{figure}

\begin{figure*}
\centering
\centerline{
    \includegraphics{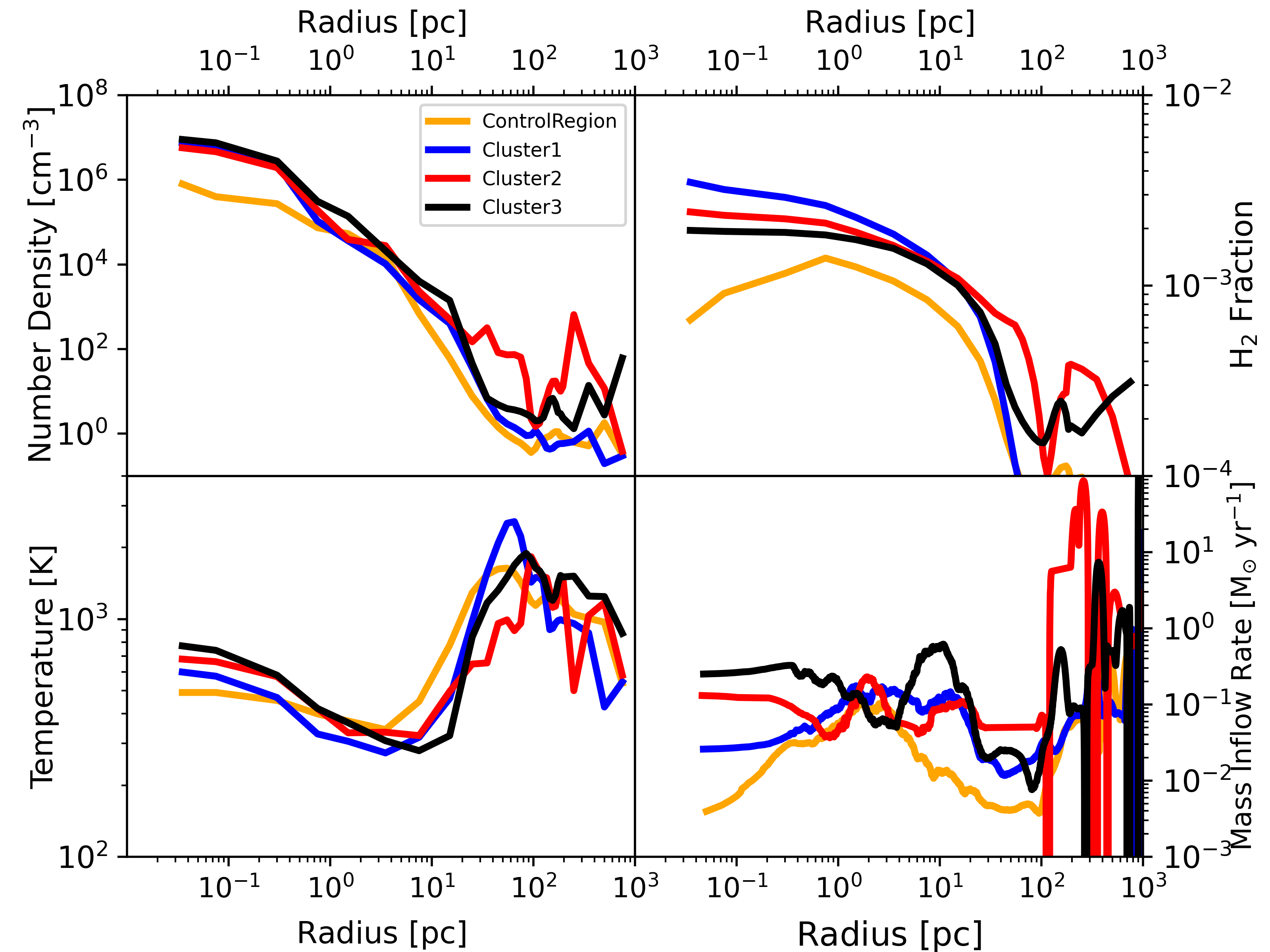}}
\caption{
  Radial profiles for the four haloes identified as undergoing first gravitational collapse -
  one from each of the ControlRegion, Cluster1, Cluster2 and Cluster3. Starting in the lower left panel and moving
  clockwise we show the gas temperature, gas number density, \molH fractions and the mass accretion
  rates into the halo centre. The virial temperature of each halo is approximately 2000 K well
  below the threshold for \lya cooling meaning that cooling is dominated by \molH emission line
  cooling. Temperatures in the centre of each halo are approximately
  700 K. Gas number densities follow an approximately isothermal profile with \molH fractions
  consistent with their equilibrium value of $10^{-3}$. The accretion rates into the halo centres
  are remarkably high with values close to, or in excess of, 0.1 \msolaryr down to small radii.
  Such high accretion rates in the halo centre
  indicate that stars with ``super-massive'' characteristics may initially form
  \cite[e.g.][]{Woods_2018}. The virial radius for the four regions are; ControlRegion : 45 pc,
  Cluster1 : 55pc, Cluster2 : 105pc, Cluster3 : 105pc.
}
\label{Fig:MultiPlot}

\end{figure*}

\begin{figure}
\centering
\centerline{
  \includegraphics[width=9.0cm]{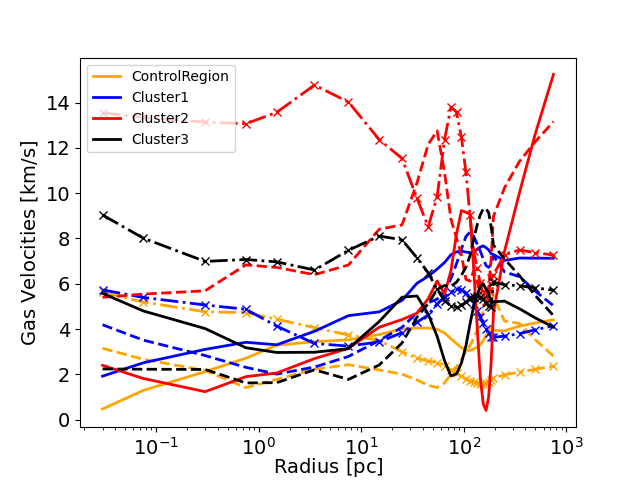}
}
\caption{Gas velocities from the first halo to collapse in the ControlRegion (orange),
  Cluster1 (blue), Cluster2 (red) and Cluster3 (black). 
  Solid lines mark radial inflow velocities,
  dashed lines the turbulent velocities and dashdot with `x' markers
  the rotational velocities. The flow is
  almost always supersonic with radial inflow velocities
  exceeding the sound speed at almost all radii. In all cases the gas is
  rotationally supported with the rotational velocities dominating over all other
  velocity components. Turbulent velocities (dashed lines) are sub-dominant in all cases.}

\label{Fig:GasVelocities}
\end{figure}

\subsection{Stellar Mass Predictions}
\noindent We can further investigate the analytical impact of rapid assembly and to what extent it may
impact on the initial mass function (IMF) of the stars that
subsequently form by appealing to the fitting functions introduced in \cite{Hirano_2014}.
Using their equation 14:
\begin{equation} \label{Eqn:HiranoRates}
  \rm{M_{popIII}} = 100 \ \rm{M_{\odot} \Big ( \frac{\dot{M}_{virial}}{1.2 \times 10^{-3} \ M_{\odot} \ yr^{-1}}} \Big)
\end{equation}
we can calculate the expected stellar masses based on the mass inflow rates at the virial radius
of our haloes. In Figure \ref{Fig:HaloAccretionRates} we show the mass inflow rates as a function
of halo radius. The virial radius for each halo varies with the mass of each halo. To guide the eye we have coloured the
range of virial radii in grey - the median value for the virial radius of all haloes is
approximately 100 pc and the average mass inflow rate within this region is $\dot{M} \sim 0.2$ \msolaryrc. \\
\indent Now using the fitting function as given in \cite{Hirano_2014} we find, given these
high mass inflow rates, predicted stellar masses in excess of $10^4$ \msolar and
possibly even exceeding $10^5$ \msolar for some haloes.
In Figure \ref{Fig:HiranoRates} we show the predicted
stellar masses for each halo as a function of halo mass. The fact that many haloes are growing
at very high rates means that Eqn \ref{Eqn:HiranoRates} predicts a large spread in final stellar
masses. However, these results should be treated with caution and do not necessarily account
for fragmentation in these haloes where bound clumps may fragment into smaller mass proto-stars
or fluctuations in the mass inflow rate -
even though this fitting function was calibrated against mini-haloes in the same mass range
(albeit ones not undergoing rapid assembly). The open circles in Figure \ref{Fig:HiranoRates}
refer again to the haloes in which star formation has occurred. For these haloes the
fitting function predicts stellar masses in range $10^3$ \msolar up to $10^5$ \msolarc.
Within an order of magnitude these values are indeed correct and capture the top-heavy nature of the
stellar masses formed. The actual stellar masses that are found in our simulations
(see \S \ref{Sec:Onset}) are however somewhat lower. Nonetheless, the fitting function does point to a significantly more
top-heavy IMF in these rare haloes forming in overdense peaks - consistent with recent results
from \cite{Regan_2020b} and \cite{Latif_2022}. Note also that the orange points (ControlRegion)
are at least an order of magnitude below the points from Cluster2 and Cluster3.\\


\begin{figure}
  \centerline{
    \includegraphics[width=9cm]{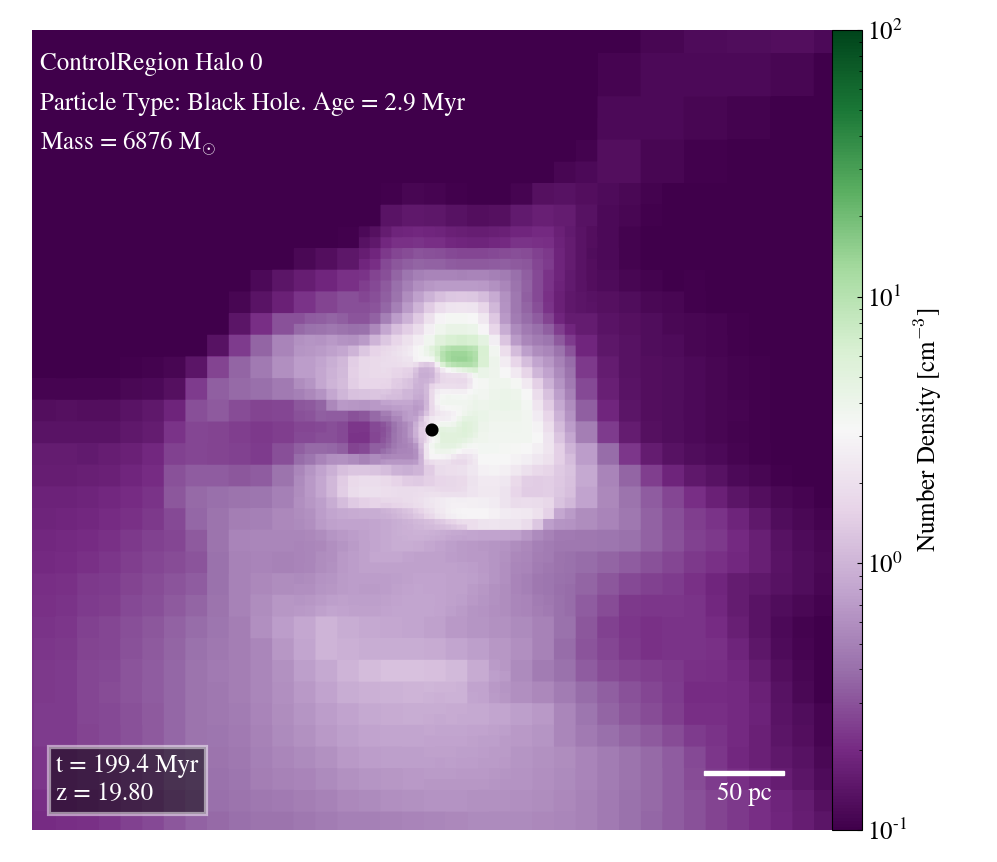}
  }
  \caption[]{\label{Fig:ControlRegion_FirstStar} The first halo to collapse and form a star in the ControlRegion
    forms a SMS initally due to the initial large mass inflow onto the stellar surface. This large mass inflow quickly grows the
    star to approximately 6000 \msolarc. The star's short lifetime (500 kyr) means that it quickly collapses into a black hole.
    The black hole sits at the centre of the halo but is currently quiescent surrounded by relatively low density gas.
}
\end{figure}


\begin{figure*}
  \centerline{
    \includegraphics[width=9cm]{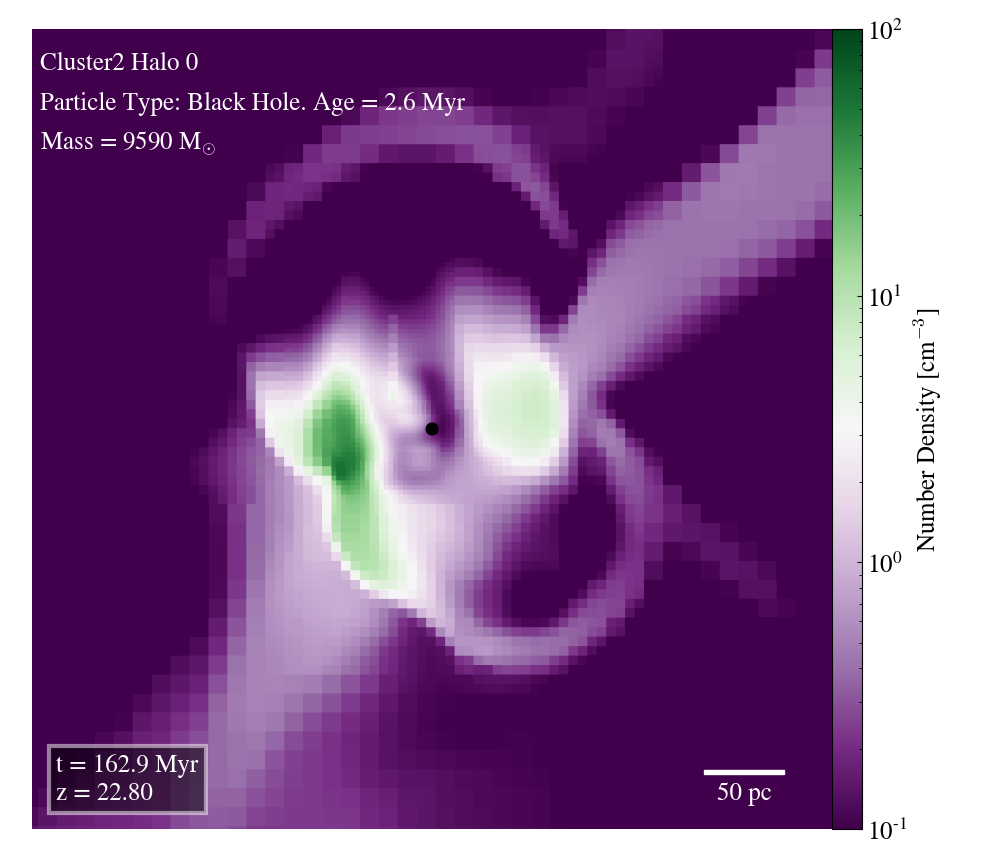}
    \includegraphics[width=9cm]{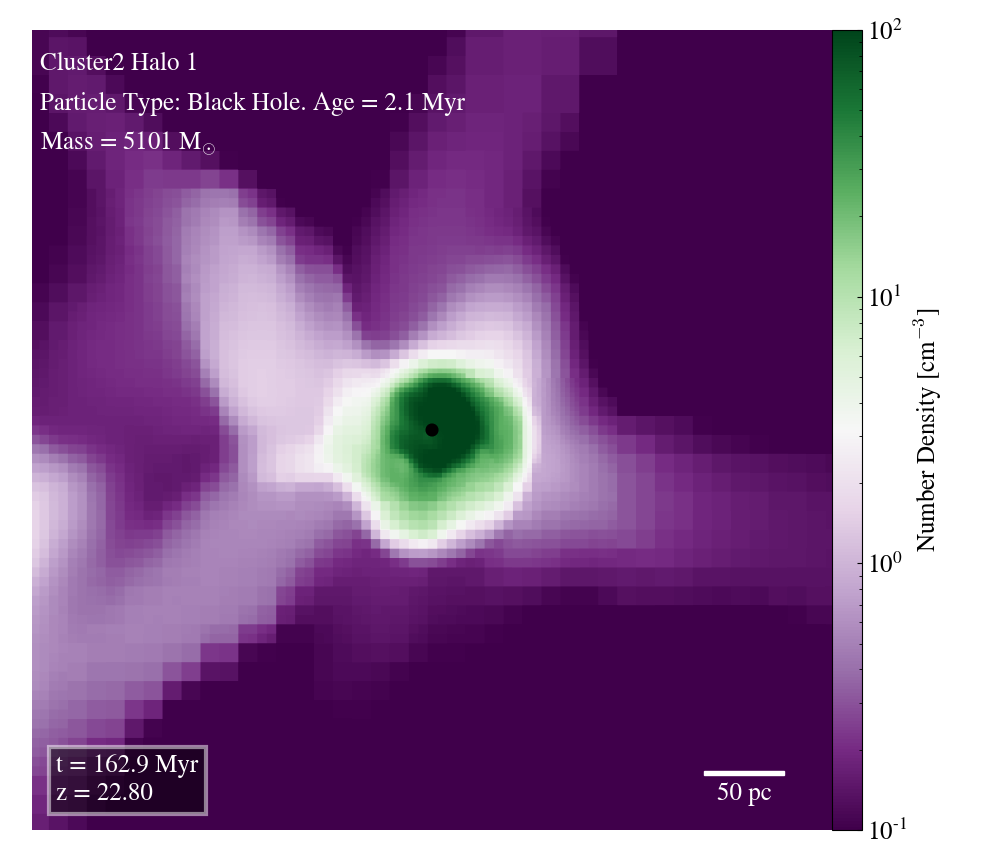}
  }
  \caption[]{\label{Fig:Cluster2_FirstStar} Two stars form in two separate haloes in Cluster 2 within the
    2.6 Myr following the formation of the first star (left panel). Both stars accrete extremely rapidly and
    using the fitting formula from \cite{Schaerer_2002} the lifetimes of these massive PopIII stars are
    300,000 and 600,000 years respectively. Following their stellar lifetimes both stars directly
    collapse into massive black holes with masses of 9590 \msolar and 5101 \msolar respectively.
}
\end{figure*}

\subsection{Radial Profiles of the Star Forming Haloes}

\noindent In Figure \ref{Fig:MultiPlot} we plot the radial profiles of the first halo to
undergo gravitational collapse in each of the ControlRegion, Cluster1, Cluster2 and Cluster3.
In the panels, starting in the bottom left and moving clockwise, we plot the temperature, number
density, \molH fraction and the mass inflow into the halo centre. The profiles of each cluster are
broadly similar. The temperature at the centre of each halo is between 600 K and 800 K typical of
galaxies hosting the formation of metal-free stars with virial temperatures of close to 2000 K.
The density profile of each halo is roughly isothermal with some flattening of the profile
towards the halo centre as the gas becomes pressure supported. \molH fractions saturate at
approximately $2 \times 10^{-3}$. The mass inflow rates into the centre of each halo are remarkably
high however and are much higher than expected for a typical PopIII forming halo \citep{Hirano_2014}.
The mass inflow rates into the centre of each halo are in the realm of that expected for supermassive
star formation \citep{Haemmerle_2018}. Although, as we will see and as was found in
\cite{Regan_2020b}, such high mass inflow rates are difficult to sustain (as stellar accretion rates)
for the entire (pre-main sequence) evolution of the star.\\
\indent A recent study by \cite{Latif_2022} (L22) found a rare halo collapsing at the nexus of a
number of converging cold flows at high redshift ($\rm{z} \sim 26$). For their halo they found that
the pressure support from turbulent velocities significantly exceeded the gravitational compression.
This allowed the halo to delay collapse until the halo had a mass of $\rm{M_{halo}} =
4 \times 10^7$ \msolarc. This is similar, albeit a factor of a few, more massive than our most
massive halo. In their study they found turbulent velocities of order a few tens of \kms. \\
\indent In Figure \ref{Fig:GasVelocities} we show
the radial velocity (solid lines), turbulent velocities (dashed lines), sound speed (dotted lines)
and rotational velocities (dashdot with 'x' marker) at the same epoch as in Figure \ref{Fig:MultiPlot}.
The rotational velocities are calculated by taking the moments of the inertia tensor of the gas
at varying radii from the centre \citep[e.g.][]{Regan_2009} with the turbulent velocities calculated
by subtracting both the bulk and radial velocities away from the raw cell velocities.
This likely gives an upper limit to the level of turbulent velocities in the halo. 
In all cases the rotational velocities dominate, indicating gas that is strongly rotationally
supported. Similarly to L22 we see evidence of supersonic radial inflows but our turbulent velocities
are approximately a factor of between three and five times lower than the turbulent velocities found
in L22. However, the haloes we show here are merely the first to collapse and hence not
representative of the rarer haloes which may be required for turbulent velocity support to take hold.
Across the mass spectrum of all haloes sampled we see strong evidence for rapid assembly being
the primary driver for delaying gravitational collapse and allowing
for the build-up of more massive haloes (and hence larger accretion rates onto the halo centres). It is
likely, and indeed L22 reach this broad conclusion, that their halo is a rarer manifestation of the
rapid assembly process with rapid global inflows driving the formation of massive stars in (rare)
massive haloes. \\
\indent In summary, for the first haloes to collapse in our sample we observe rotationally
supported gas dominating which is also likely to impact the mass of the stars formed unless the
rotational support can be transported away efficiently.


\begin{figure*}
  \centerline{
    \includegraphics[width=9cm]{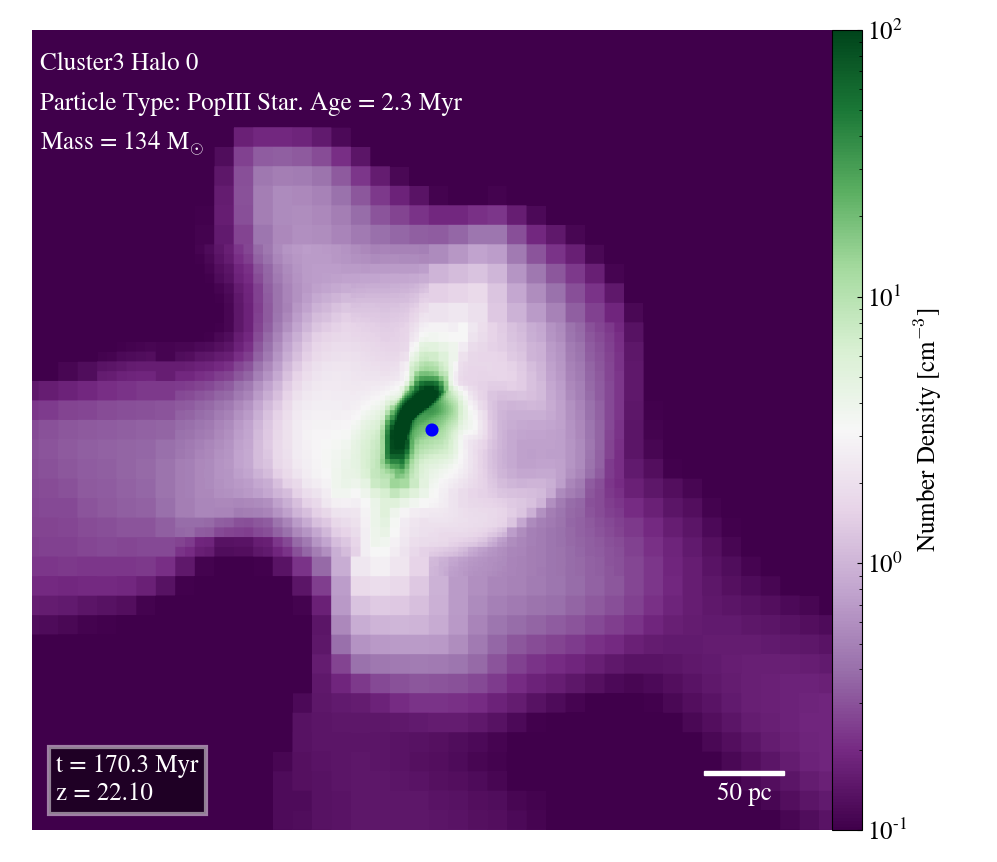}
    \includegraphics[width=9cm]{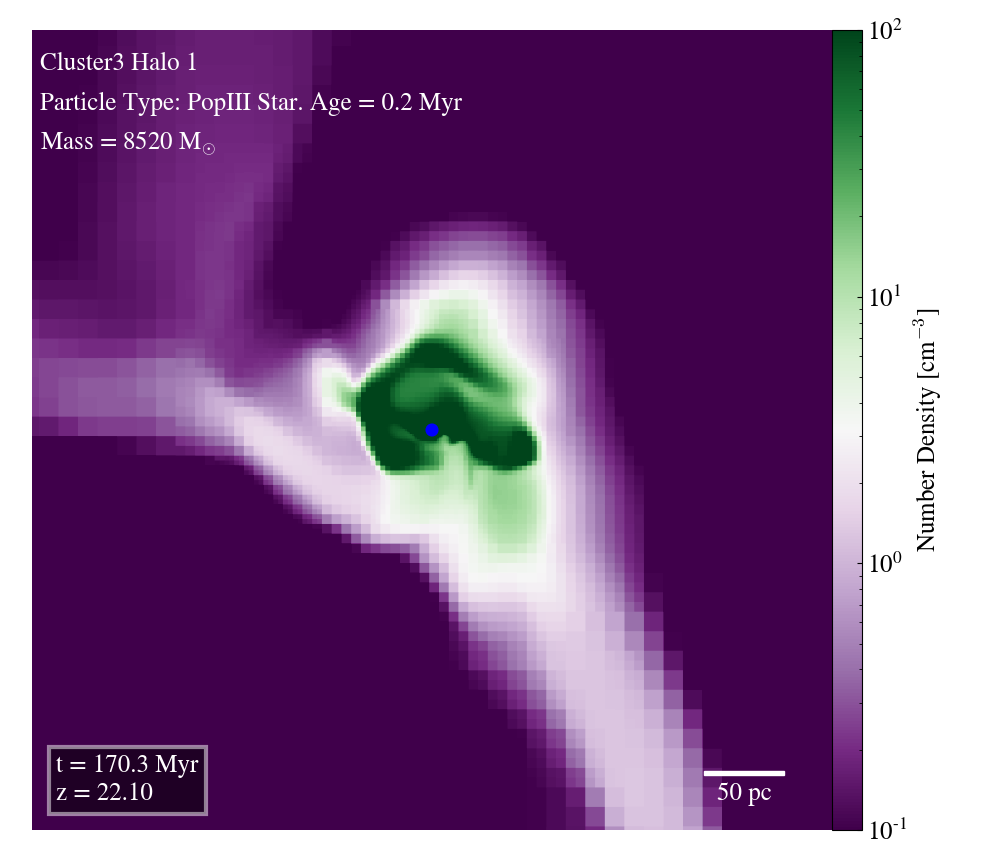}
  }
  \caption[]{\label{Fig:Cluster3_FirstStar} Similar to Cluster 2 with two stars in two separate haloes
    within the Cluster 3 environment. The first star to form in this case was a PopIII star with an (IMF
    sampled) mass of 134 \msolarc. In our simulations PopIII stars do not accrete as their spatial scale
    is below our resolution scale. The lifetime of a star of this mass is calculated to be 2.6 Myr and
    has therefore not yet collapsed into a black hole. The other star is also a massive PopIII star with
    a current age of 0.2 Myr. It's calculated lifetime is 0.6 Myr (due to it's extreme mass) and it will
    collapse into a massive black hole at that time. 
}
\end{figure*}

%
\begin{figure*}
  \centerline{
    \includegraphics[width=9cm]{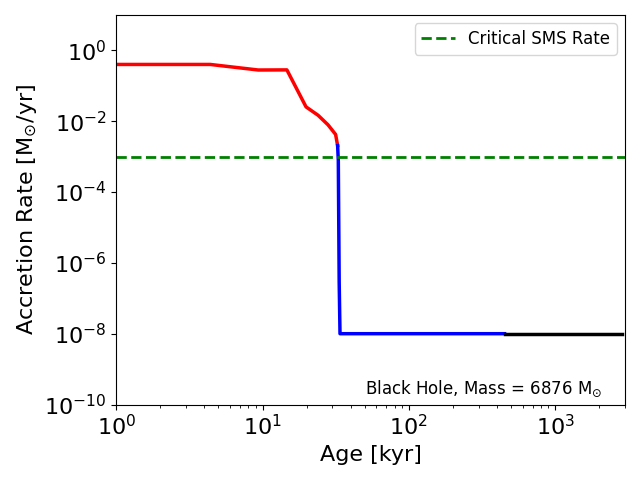}
    \includegraphics[width=9cm]{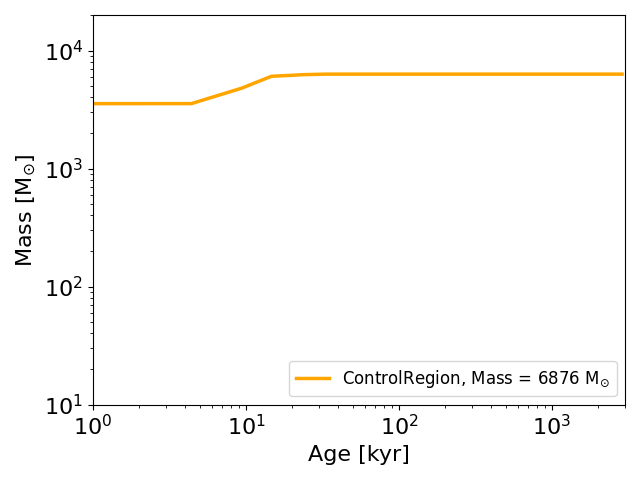}
  }
  \caption[]{\label{Fig:ControlRegion_AccretionRates}
    In the left hand panel we show the accretion rate onto the first star to form in
    the ControlRegion. The star initially and rapidly accretes
    matter onto the stellar surface at rates significantly exceeding the critical rate. This causes
    the star to move to the red, puff up, and become a SMS. However, the high accretion rates are
    not maintained and after approximately 50,000 years the star transitions to the blue and
    become a massive PopIII star. The star subsequently collapses into a massive black hole.
    Right Panel: The mass history of the star/MBH. The rapid increase in mass at early times
    is the dominant driver in determining the final
    mass of the star.
}
\end{figure*}

%
\begin{figure*}
  \centerline{
    \includegraphics[width=9cm]{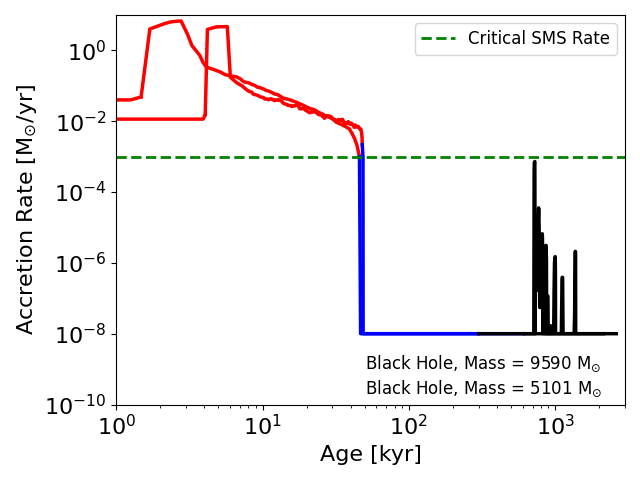}
    \includegraphics[width=9cm]{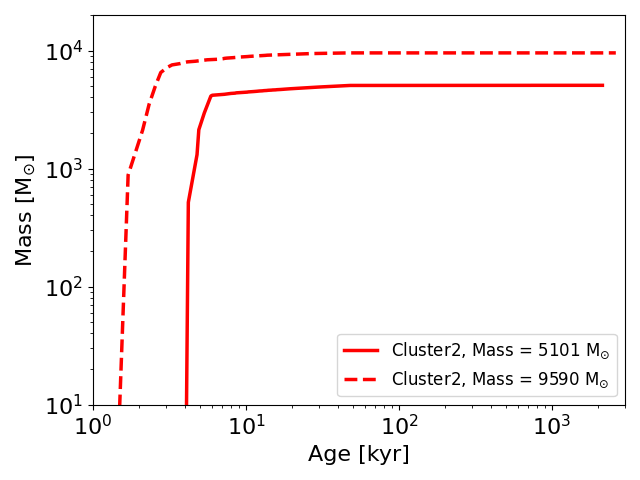}
  }
  \caption[]{\label{Fig:Cluster2_AccretionRates}
    In the left hand panel we show the accretion rates for the two objects that have formed in
    Cluster2. The evolution of both stars follows a similar trajectory to the star that forms in
    the ControlRegion. Both stars collapse into MBHs at the end of the stellar lifetime.
    One of the black holes exhibits intermittent accretion approximately 1 Myr after formation
    as it passes through some dense material (see right hand panel of
    Figure \ref{Fig:Cluster2_FirstStar}). However, the accretion is relatively short lived and
    has little or no impact on the mass of the black hole. Right Panel: The mass history of both
    objects. Again the rapid increase in mass at early times is the dominant driver in determining the final
    mass of the star. The small accretion activity seen at approximately 1 Myr in the left panel
    has no appreciable effect on the mass of the black hole. 
}
\end{figure*}

%
\begin{figure*}
  \centerline{
    \includegraphics[width=9cm]{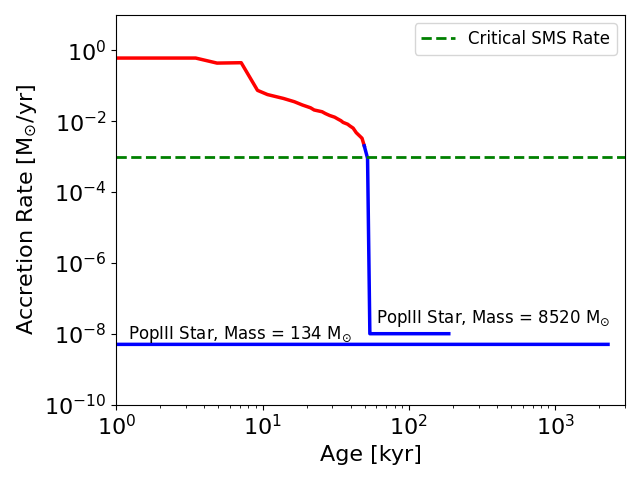}
    \includegraphics[width=9cm]{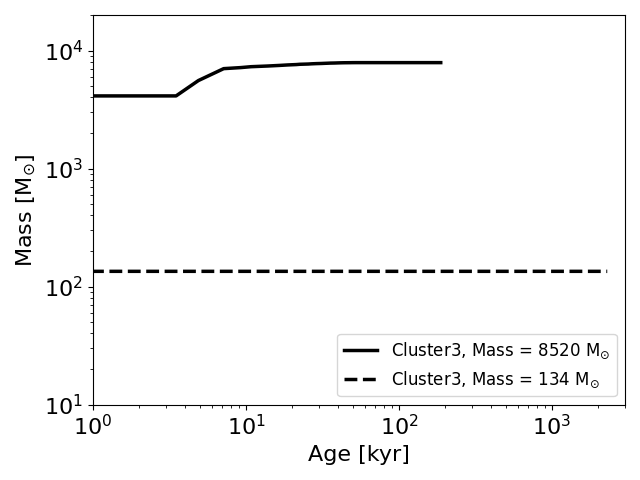}
  }
  \caption[]{\label{Fig:Cluster3_AccretionRates}
    In a very similar pattern to Cluster2, in the left hand panel we show the accretion rates for the two objects that have formed in
    Cluster3. In the case of Cluster3 however the first star forms as a PopIII star (with a mass of 134 \msolarc) while the
    second star forms as a SMS. The SMS subsequently contacts to the main sequence and evolves into a massive PopIII star
    with a mass of 8520 \msolarc. Both stars are currently still in their stellar phases. 
    Right Panel: The mass history of both stars - the first star to form (a PopIII star) does not accrete and has
    the same mass throughout. The second star (initially a SMS) grows rapidly in mass before contracting to the
    main sequence. 
}
\end{figure*}

\section{The onset of Star Formation} \label{Sec:Onset}
\noindent Having found that Cluster2 and Cluster3 show a preponderance of rapidly accreting haloes we
proceed to evolve Cluster 2 and Cluster3 beyond the initial phase of gravitational collapse.
This is also necessary as we need to focus our computational resources. We additionally continue to
evolve the ControlRegion as a comparison against a `normal' region. 
Cluster 2 is the first halo to collapse and undergo star formation at z = 23.06 followed by
Cluster3 at z = 22.31. The ControlRegion meanwhile forms its first star at z = 20.00. In
Figure \ref{Fig:ControlRegion_FirstStar}, Figure \ref{Fig:Cluster2_FirstStar} and 
Figure \ref{Fig:Cluster3_FirstStar} we show visualisations of each of the star forming
haloes. The visualisation for the ControlRegion is created at z = 19.80 at which point the
massive PopIII star has collapsed into a MBH. It's current age is 2.9 Myr. The star forming
haloes in Cluster2, of which there are two,
are separated by 33 kpc (physical). The visualisations (centred on the stars in each case)
are both created at z = 22.8 which is more than 2 Myr after the formation of the first star.
As can be seen from the visualisations both stars are embedded within an overdensity. 
The results for the star forming haloes in Cluster3 are broadly similar (see Figure \ref{Fig:Cluster3_FirstStar})
with again two different star forming haloes emerging. In this case however the separation between the haloes
is significantly less with a separation of less than 1 kpc. While the separation both in time and
space between the haloes is (marginally) consistent with the synchronised pair scenario the haloes
are both too small and there is no background LW radiation field to support the mechanism
\citep{Dijkstra_2008, Visbal_2014b, Regan_2017}. \\
\indent In Figure \ref{Fig:ControlRegion_AccretionRates}, Figure \ref{Fig:Cluster2_AccretionRates}
and Figure \ref{Fig:Cluster3_AccretionRates} we show the accretion rates and mass growth history of each of the
five stars to form as well as their transitions into massive PopIII and black holes where appropriate.
In Figure \ref{Fig:ControlRegion_AccretionRates} (ControlRegion) we can see that the initial (merger triggered)
star formation results in a SMS with growth rates exceeding the critical value. After approximately 50 kyr
the growth rate of the star drops off as the gas is accreted by the star. At this point the star has grown to a
mass of 6876 \msolarc. In Figure \ref{Fig:Cluster2_AccretionRates} we show the accretion rates and masses of
the two stars from Cluster2. 
Both stars initially accrete extremely rapidly with mass accretion rates well in excess of
$10^{-3}$ \msolaryr which we use here as the critical rate for SMS formation \citep{Haemmerle_2018}.
However, in both cases this rapid accretion is short-lived and ceases a few tens of kiloyears
after formation. At this point both stars contract towards the main sequence and become massive
PopIII stars - similar to the evolution seen in the ControlRegion halo. The large masses of these objects
also means that their lifetimes are extremely short. We calculate the lifetimes of these massive PopIII stars by fitting
to the functional form given by \cite{Schaerer_2002}. Given their masses these massive PopIII stars
directly collapse into massive black holes after a few hundred kiloyears. Similar to the case of when
the stars were SMSs the accretion rates remain extremely low following their transition into MBHs.
One of the MBHs (the MBH with a mass of 5101 \msolarc) does accrete intermittently with rates up to
$10^{-3}$ \msolaryr but these rates hardly effect the mass. The accretion does give rise to small
amounts of feedback however which we discuss in the next section. \\
\indent The first star to
form in Cluster3 forms as a PopIII star with an IMF sampled mass of 134 \msolar - it's initial
accretion rate is below that required for SMS formation and so
a PopIII star is formed from the start. PopIII stars do not accrete in our simulations and hence
this star remains a PopIII star with an unchanging mass.
The calculated age of the star is 2.6 Myr and at this stage (current age = 2.3 Myr) the star is
still in that phase. The other star to form (in a separate halo) has an initial accretion rate
exceeding the critical threshold rate and therefore forms a SMS (see Figure \ref{Fig:Cluster3_FirstStar}).
In a similar vein to the stars in the ControlRegion and Cluster2 the initial accretion rate is not maintained and once the
gas in the immediate vicinity of the SMS is accreted the accretion rate plummets and the star contracts to the main
sequence. The SMS, which is the second star to form in Cluster3, has a mass
of 8676 \msolar when it transitions to a massive PopIII star as shown in Figure
\ref{Fig:Cluster3_AccretionRates}. This star's lifetime is calculated at 0.6 Myr
(current age = 0.2 Myr) and will directly collapse into a black hole at that point. We will report
the finding of the growth and dynamics of these first MBHs in a subsequent paper.

\subsection{Merger Triggered Star Formation at High-z}
\begin{figure*}
\centering
\begin{minipage}{175mm}      \begin{center}
\centerline{
    \includegraphics[width=16.0cm, height=10cm]{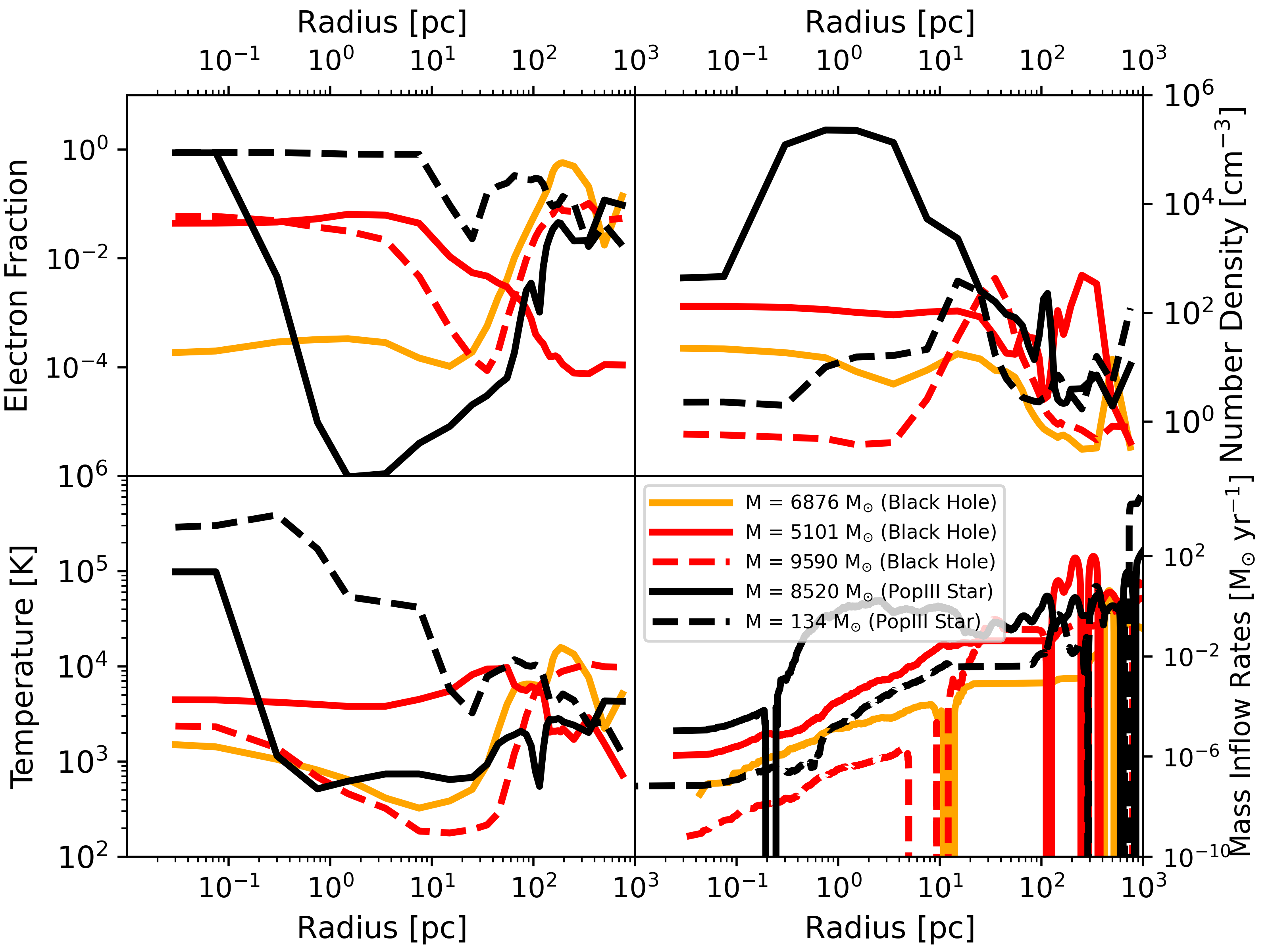}}
\caption{Radial profiles for the host halo in each cluster which hosts a star or MBH.
  Orange lines for the ControlRegion, red lines for Cluster2 and black lines for Cluster3.
  The gas surrounding the MBHs is cooler
  due to the lack of radiative feedback from the non-accreting MBHs. Electron fractions are
  very high within 0.1 pc of each object but drop off for the cases of all objects apart from the
  PopIII star with a mass of 134 \msolarc. Number densities in the host haloes are relatively low
  with the exception of the halo hosting the 8520 \msolar PopIII star. This star is surrounded, at
  a distance of less than 1 pc, by high density gas which, depending on subsequent angular momentum,
  may be accreted on the MBH that will form with a few hundred kiloyears. Mass inflow rates into
  this host halo are also favourable exceeding $10^{-2}$ \msolaryr within 1 pc of the halo centre. }
\label{Fig:StarFormationRadialProfiles}
\end{center} \end{minipage}
\end{figure*}

\begin{figure*}
  \centering
  \includegraphics[width=8.5cm, height=6cm]{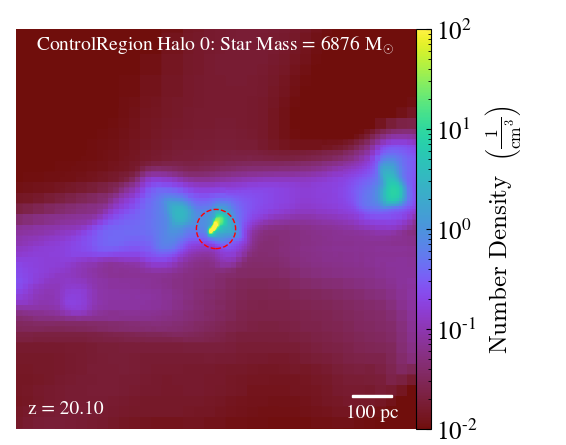}
  \includegraphics[width=8.5cm, height=6cm]{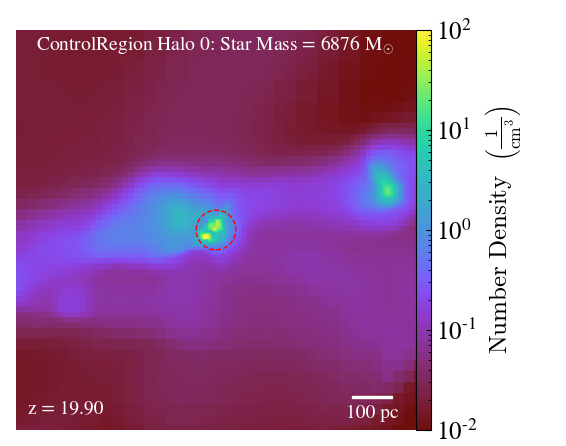}\newline
  \includegraphics[width=8.5cm, height=6cm]{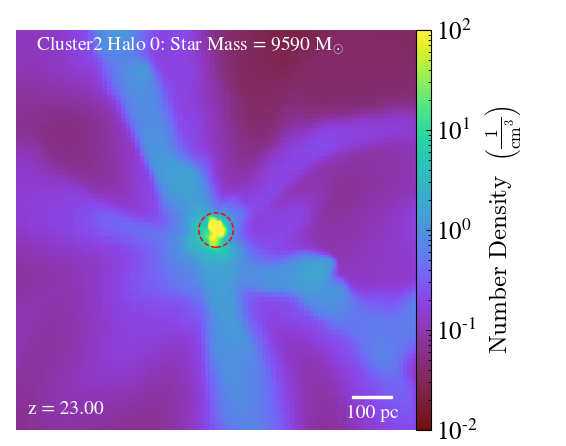}
  \includegraphics[width=8.5cm, height=6cm]{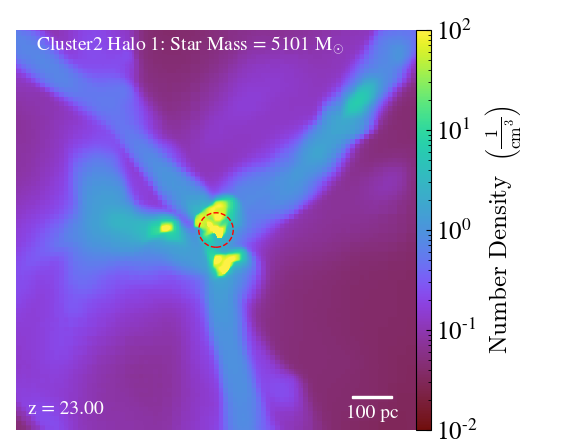} \newline
  \includegraphics[width=8.5cm, height=6cm]{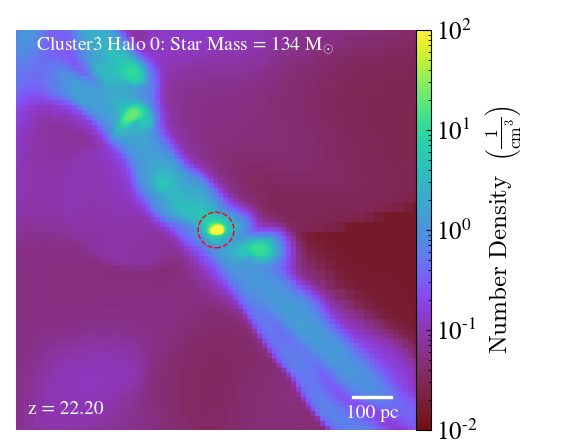}
  \includegraphics[width=8.5cm, height=6cm]{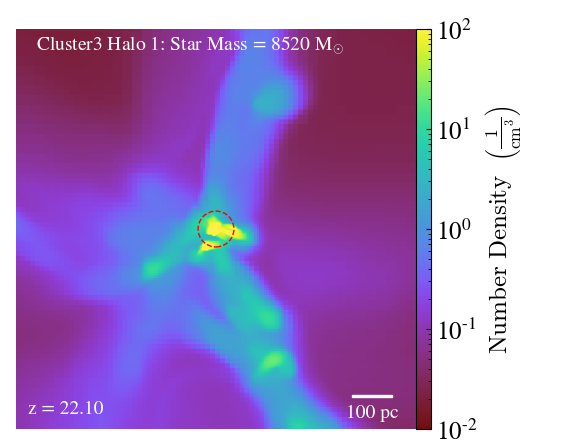}
  \caption[]{\label{Fig:Filaments} The large scale environment surrounding each of the star forming haloes at the output redshift closest to the onset
    of star formation. The red dashed circle in each case marks the
    star forming halo. \textit{Top row:} The first row shows the conditions operating around the ControlRegion halo. There is only one star forming halo in the ControlRegion and so we show two outputs.
    Between $z \sim 20.10$ and $z \sim 19.90$ a merger
    occurs between the most massive (circled) halo and a smaller halo to its immediate left. The merger actually causes the original halo to be disrupted but
    in doing so also triggers star formation (at $z \sim 20$). \textit{Middle row:} The second row shows the conditions around the two star forming haloes in Cluster2.
    Each halo is positioned at the nexus of intersecting filaments. Halo 0 (middle left panel) forms a star due to triggered star formation while
    Halo 1 (middle right panel) experiences more modest growth and forms a star prior to the halo reaching the atomic cooling limit.
    \textit{Bottom row:} The third row shows the conditions around the two star forming haloes in Cluster3. In Halo 0 (bottom left panel), which is located,
    mid-way along a filament a 134 \msolar star forms. In Halo 1 (bottom right panel) the star forming halo is located again at the nexus of a number
    of filaments which delays star formation until the halo approaches the atomic cooling limit.}
\end{figure*}

\noindent The overdense nature of Cluster2 and Cluster3 combined with the relatively youthful age of the Universe
at this epoch means that mergers are common (the Universe at $z \sim 20$ is approximately
10,000 times more dense than today's Universe). Examining again
Figure \ref{Fig:MassRedshift} we can see that the star forming halo in the ControlRegion,
one of the star forming haloes in Cluster2 and one of the star forming haloes in Cluster3
show steep increases in mass either just after star formation or just prior to star
formation (the point of star formation is marked by the star symbol).
In each of these cases a major merger takes place resulting in the rapid flow of gas to the halo centre.
The mass inflow rates due to these mergers vary between $10^{-3}$ \msolaryr and 1 \msolaryr (see Figure \ref{Fig:HaloAccretionRates}
and Figure \ref{Fig:MultiPlot}). The rapid inflow of gas as a result of the merger is however time limited and
in each case, across all three of these haloes, lasts for approximately 50 kyr. The combination of this timescale
and the inflow rates leads to final stellar masses of a few thousand solar masses for two of the three haloes. The third halo
(Cluster3) which shows a steep growth profile actually forms a star prior to the merger (see below). \\
\indent Following the merger and the subsequent star formation we see the
inflow rates subside (see lower right panel of Figure \ref{Fig:StarFormationRadialProfiles}) and fall to values as
low as $\sim 10^{-8}$ \msolaryrc. For the other two star forming haloes shown in Figure \ref{Fig:MassRedshift},
which show more modest growth, albeit to higher halo masses, as opposed to the
steep ascent, the resulting stellar masses are actually quite similar (these two haloes form
stars with masses of 5101 \msolar and 8520 \msolar respectively). These two, more modestly growing, haloes are
positioned at the nexus of multiple filaments (see Figure \ref{Fig:Filaments}) and grow to close to the atomic cooling threshold before star formation
commences. We may be witnessing a bi-modality emerging. \\
\indent Rapid growth drives very massive star formation in all cases. If a major merger occurs (which accelerates the process) then a
massive star forms from the rapid inflow due to the merger. In the absence of a major merger but in the presence of multiple
minor mergers rapid assembly dominates, preventing normal PopIII star formation, resulting nonetheless in a massive star. \\
\indent In any case for, at least some of, the first haloes to form, star formation is driven by 
\textit{merger triggered star formation}. A major merger of mini-haloes results in inflow rates sufficient to
drive the formation of a SMS but not sufficient to stop the star from returning to the main sequence after a few tens of kiloyears.
The halo with the 134 \msolar PopIII star is more standard among the five star forming haloes (bottom left panel of Figure \ref{Fig:Filaments}).
Looking again at Figure \ref{Fig:MassRedshift} we can clearly see that this
halo also suffers a major merger at $z \sim 22.2$. However, in contrast to at least three of the other halos this
halo forms along a filament as opposed to at the junction of several filaments. While this halo clearly experiences a merger, star formation occurs
prior to the merger and as a result the mass inflow rates do not exist to drive SMS formation in the first instance.
Instead in this case a PopIII star forms with a (stocastically calculated) mass of 134 \msolarc. \\
\indent In summary, merger triggered star formation drives SMS star formation in two out of our
five haloes, in a further two haloes rapid assembly grows the host halo to close to the atomic cooling threshold before star formation occurs. In the final halo a more normal PopIII star forms just prior to a major merger. There are strong parallels with our
merger triggered star formation and the proposed process for SMBH formation outlined by \cite{Mayer_2010}. \cite{Mayer_2010} proposed a scenario where
a major merger between two massive galaxies (at $z \sim 10$)  results in the direct formation of a SMBH. In our case we observe a similar process,
acting on smaller mass scales, resulting in the formation of a SMS.

\subsection{Profiling the Stellar Environment}
\noindent In Figure \ref{Fig:StarFormationRadialProfiles} we show the radial profiles for the temperature,
electron fraction, number density and mass inflow rates within the host halo of each star forming halo in each
cluster (orange for the ControlRegion, red for Cluster2 and black for Cluster3). Of the five stars which have
formed in the simulations three have since collapsed into MBHs with one residing in the ControlRegion with a
mass of 6876 \msolar and two others in 
Cluster 2 with masses of 9590 \msolar and 5101 \msolar respectively. 
All three MBHs are currently quiescent although the lighter MBH in Cluster2 has experienced some modest
accretion over last few hundred kiloyears. This may not be surprising since the gas surrounding the
lighter MBH in Cluster2 is relatively cold and much denser
compared to the gas surrounding the more massive MBH. The gas is in both cases strongly ionised with
electron fractions close to 0.1. This
is likely due to the recent stellar feedback which would have ionised the gas in the immediate
vicinity of the stars (prior to the black hole phase at which point feedback ceased
as accretion is essentially nonexistent). \\
\indent The most strongly ionised gas is that surrounding the stars
in Cluster3 which are both still in their stellar phase.
For both of these stars the gas is highly ionised (at least in the immediate vicinity of the star)
due to the radiative feedback from the PopIII star with a mass of 134 \msolar
and the thermal feedback from the massive PopIII star with a mass of 8520 \msolarc. We implement
thermal feedback for all massive PopIII stars with masses
in excess of 500 \msolarc. As noted in \S \ref{Sec:Feedback} we do this as the computational cost of
radiative feedback for such massive stars makes tracking that radiation spectrum computationally
intractable and instead we turn to thermal feedback. We do not expect this to adversely effect our
results in general since radiation even from such massive stars is not expected to propagate beyond the
immediate vicinity of these stars \citep{Woods_2021, Jaura_2022}. The ionised fraction for the
lighter PopIII star does extend beyond the vicinity of the star in this halo due to the
relative underdensity of gas in the halo and the relatively low mass of the halo. \\
\indent Infact the density of gas surrounding most of the stars and MBHs is comparatively low
(less than $10^2 \ \rm{cm^{-3}}$ for four of the five objects). Only for the (non-accreting)
8520 \msolar star does the number density exceed $10^2 \ \rm{cm^{-3}}$. This will mean that the
black hole that forms at the end of this star's stellar life (in approximately 0.4 Myr) will have
the potential to accrete a substantial mass of gas. This supposition is further supported by
examining the lower right panel of Figure \ref{Fig:StarFormationRadialProfiles}. In
that panel we can see that the mass inflow rates for this halo are the highest of all four
haloes with gas inflow rates of at least $\dot{M} \sim 10^{-3}$ \msolaryr extending
  all the way to the centre of the halo. Each of the other haloes now show
  relatively weak mass inflow rates which supports the lacklustre accretion rates seen for the
  two black holes in Cluster2.

\subsection{Numerical Caveats} \label{Sec:Caveats}
\noindent As with any numerical simulation attempting to model complex processes such as star
formation there are inevitable caveats. The primary caveat in these simulations is resolution.
Our maximum resolution here is set by the size of our smallest grid cell which is approximately
0.05 pc (physical). This is significantly above the scale at which star formation takes place and
at densities many orders of magnitude below the actual formation threshold. For example
\cite{Prole_2022} see no numerical convergence in their PopIII star formation simulations
even at densities up to $n \sim 10^{20}$ cm$^{-3}$. The results of \cite{Prole_2022} suggest that
our results here may be missing significant fragmentation of the central core and that many
PopIII stars may form in these first mini-haloes in contrast to the one SMS we see in our haloes.\\
\indent However, this must be weighed against  the fact that these haloes here experience mass inflow rates
many orders of magnitude higher than average and the fact that our sink particle calculated masses
are below those predicted by the fitting functions of \cite{Hirano_2014} for example. What is
not in doubt is that we are modelling rare haloes, undergoing rapid assembly with large mass inflows, and
that they likely produce very massive stars or dense stellar clusters containing an
equivalent mass in stars \citep{Prole_2022}. Further zoom-in simulations at significantly enhanced
spatial and mass resolution will be required in the future to probe the different outcomes in greater
detail.

\section{Discussion and Conclusions} \label{Sec:Discussion}
\noindent Here we introduce the \blackdemon simulations. The rationale driving these simulations is to
investigate the formation of MBH seeds in overdense regions of the
embryonic universe in a self-consistent manner. Three distinct clusters (each 1 \mpch comoving
on the side) are identified and resolved with particle masses of
$\rm{M_{part}} \sim 2.3 \times 10^3$ \msolar and maximum spatial resolutions of
$\Delta x \sim 0.05$ pc (physical at z = 20). We additionally evolve a ControlRegion within a mean
density region of the parent box for comparison. We find that of the
three (overdense) clusters modelled, Cluster2 and Cluster3 have a significant fraction of
haloes growing at rates that exceed the critical growth rate and show the largest deviations from
a predicted Press-Schecter halo mass function. \\
\indent As a result the cooling times of the gas for a significant number of haloes in
Cluster2 and Cluster3 are longer than the analytically predicted values, these
longer cooling times allow larger Jeans masses to develop lending support to the
premise that these haloes will produce more massive stars. We find that the
first halo to collapse (at $z = 23.06$) has a mass of $1.04 \times 10^7$
\msolar when it first undergoes gravitational instability. This mass is just short
of the atomic cooling threshold at this redshift ($M_{\rm{atm}} \sim 1.5 \times 10^7$
\msolar at $z \sim 23$). \\
\indent We continue the evolution of the ControlRegion, Cluster2 and Cluster3 beyond the formation
of the first star in each case and report the results approximately 2 Myr after first star formation in each
region. After 2 Myr two MBHs reside in two separate haloes in
Cluster2 with masses of 9590 \msolar and 5101 \msolar respectively. In both cases initially
a SMS formed but in both cases the SMS contracted to the main sequence as critical accretion was not
maintained. Both MBHs are currently quiescent. Cluster3
shows a broadly similar result. In Cluster3, two PopIII stars reside with masses of 134 \msolar and
8520 \msolar respectively. Both objects are currently still in their stellar phases and have not yet
collapsed to form MBHs. Interestingly the ControlRegion also forms a SMS in the first halo to collapse
in that region. As discussed in \S \ref{Sec:Caveats} higher resolution simulations of these environments
will ideally be required to resolve the formation of a molecular core \citep[e.g.][]{Prole_2022} and to probe
fragmentation below our current resolution limit. \\
\indent In investigating the driver of the massive star formation in each halo we found that
major merger triggered star formation occured in two haloes (one in the ControlRegion and
one in Cluster2). In another two cases (one in Cluster2 and one in Cluster3)
massive star formation occured following rapid assembly up to just short of the
atomic cooling limit in haloes at the junction of multiple filaments. In the fifth and final case a
`normal' PopII star formed.\\
\indent \textit{What then can we extract from the initial results of these simulations?} The first
result is that haloes evolving inside an overdense region, in the case of Cluster2 and Cluster3 a
4-sigma fluctuation, experience large mass infall rates driven by the numerous minor mergers they
experience. Secondly, the minor mergers in many cases result in dynamical heating mitigating against early PopIII
star formation allowing the build up of larger halo masses. Thirdly, major mergers once the masses of the
haloes approach a few times $10^6$ \msolar can also drive SMS star formation. The mergers at this point push the gas past the point
of gravitational instability and result in massive star formation. The large gas inflows from the merger are of the order of 0.01 -
1 \msolaryr in the newly formed halo centre and trigger SMS formation. These rapid mass inflows are however not sustained and
after approximately 50 kyr the mass inflow rates subside and the SMS contacts back to the main sequence forming a massive PopIII
star with a mass in the range 5,000 - 10,000 \msolarc. Given the overdense environment further major mergers may however be expected
which will replenish the gas reservoir at the centre of these haloes. At the time of reporting three of the first stars to form
have transitioned into MBHs with masses of 5101 \msolarc, 6876 \msolar \& 9590 \msolarc.
The MBHs are currently quiescent (though one showed short-lived accretion)
and must await further gas inflows before being able to grow further. The two other stars remain in their PopIII stellar phase with masses of
134 \msolar and 8520 \msolar respectively. \\
\indent What we are probing here is the high end tail of the PopIII mass spectrum. These
are stars forming in rare, over-dense, environments experiencing mass infall rates
10's to 100's of times the average values. The star formation is being driven by the
mergers of building block sub-embryonic galaxies - a regime previously almost universally unexplored.
The first generation of results from \cite{Bromm_2002} and \cite{Abel_2002} predicted that the first stars would be very
massive with masses close to 1000 \msolarc. Subsequent and more
detailed calculations at higher resolution and updated chemistry solvers showed that
fragmentation of the initial gas cloud reduced the final masses considerably
\citep[e.g.][]{Clark_2008, Turk_2009}. However, it now appears that by probing the high-end tail
of the distribution that the very first stars may indeed have been exceptionally massive with final
masses well in excess of $10^3$ \msolarc. \\

\section*{Acknowledgments}

\noindent JR acknowledges support from the Royal Society and Science Foundation Ireland under
grant number URF$\backslash$R1$\backslash$191132. JR also acknowledges support from the Irish
Research Council Laureate programme under grant number IRCLA/2022/1165.
JR acknowledges that the results of this research have been achieved using both
EuroHPC JU (Karolina) and DECI resources (Barbora)
supported by the IT4Innovations National Supercomputing Center based in the Czech Republic.
JR also wishes to acknowledge the DJEI/DES/SFI/HEA Irish Centre for High-End Computing (ICHEC) for the
provision of computational facilities and support on which a portion of these simulations were run.
Additionally portions of the simulations were run on the Frontera Supercomputer based at the
Texas Advanced Computing Center, US. The freely available plotting library {\sc
matplotlib} \citep{matplotlib} was used to construct numerous plots within this
paper. Computations and analysis described in this work were performed using the
publicly-available \enzo{}\citep{Enzo_2014, Enzo_2019} and \yt{} \citep{YT} codes,
which are the product of a collaborative effort of many independent scientists
from numerous institutions around the world. Their commitment to open science
has helped make this work possible. JR finally thanks John Wise for constructive discussions and
feedback on this manuscript and in general over many years! We thank the anonymous referee for a
constructive report.

\label{lastpage}
\bibliographystyle{mn2e}
\bibliography{mybib}
\end{document}